\documentclass[10pt,a4paper,preprint, superscriptaddress]{revtex4-2}
\usepackage{float}
\usepackage{amsmath,amssymb}
\usepackage[utf8]{inputenc}
\usepackage{graphicx}
\usepackage[english]{babel}
\usepackage{pdfpages}

\makeatletter
\AtBeginDocument{\let\LS@rot\@undefined}
\makeatother

\begin{document}

\title{From Processing to Functionality: Engineering Accessible Material States in Cu-Embedded SiO$_x$ Memristive Devices}
\date{\today}
\author{Tobias Gergs}
\email[]{tobias.gergs@enas.fraunhofer.de}
\affiliation{Theoretical Electrical Engineering, Department of Electrical and Information Engineering, Kiel University, Kaiserstraße 2, 24143 Kiel, Germany}
\affiliation{Chair of Applied Electrodynamics and Plasma Technology, Faculty of Electrical Engineering and Information Technology, Ruhr University Bochum, 44780 Bochum, Germany}
\affiliation{Fraunhofer Institute for Electronic Nano Systems ENAS, 09126 Chemnitz, Germany}
\author{Rouven Lamprecht}
\affiliation{Nanoelectronics, Department of Electrical and Information Engineering, Kiel University, Kaiserstraße 2, 24143 Kiel, Germany}
\author{Sahitya Yarragolla}
\affiliation{Theoretical Electrical Engineering, Department of Electrical and Information Engineering, Kiel University, Kaiserstraße 2, 24143 Kiel, Germany}
\affiliation{Chair of Applied Electrodynamics and Plasma Technology, Faculty of Electrical Engineering and Information Technology, Ruhr University Bochum, 44780 Bochum, Germany}
\affiliation{Kiel Nano, Surface and Interface Science KiNSIS, Kiel University, Christian-Albrechts-Platz 4, 24118 Kiel, Germany}
\author{Ole Gronenberg}
\affiliation{Kiel Nano, Surface and Interface Science KiNSIS, Kiel University, Christian-Albrechts-Platz 4, 24118 Kiel, Germany}
\affiliation{Synthesis and Real Structure, Department of Materials Science, Kiel University, Kaiserstraße 2, 24143 Kiel, Germany}
\author{Luca Vialetto}
\affiliation{Theoretical Electrical Engineering, Department of Electrical and Information Engineering, Kiel University, Kaiserstraße 2, 24143 Kiel, Germany}
\affiliation{Department of Mechanical and Aerospace Engineering, University of California, Los Angeles, Los Angeles, CA 90095, United States of America}
\author{Hermann Kohlstedt}
\affiliation{Nanoelectronics, Department of Electrical and Information Engineering, Kiel University, Kaiserstraße 2, 24143 Kiel, Germany}
\affiliation{Kiel Nano, Surface and Interface Science KiNSIS, Kiel University, Christian-Albrechts-Platz 4, 24118 Kiel, Germany}
\author{Thomas Mussenbrock}
\affiliation{Chair of Applied Electrodynamics and Plasma Technology, Faculty of Electrical Engineering and Information Technology, Ruhr University Bochum, 44780 Bochum, Germany}
\author{Jan Trieschmann}
\email[]{jt@tf.uni-kiel.de}
\affiliation{Theoretical Electrical Engineering, Department of Electrical and Information Engineering, Kiel University, Kaiserstraße 2, 24143 Kiel, Germany}
\affiliation{Kiel Nano, Surface and Interface Science KiNSIS, Kiel University, Christian-Albrechts-Platz 4, 24118 Kiel, Germany}

\begin{abstract}
Resistive switching in oxide-based devices is widely governed by stochastic defect processes, yet a predictive link between fabrication conditions and functional behavior remains elusive. Here, we establish a multiscale framework connecting plasma-defined deposition conditions to macroscopic device functionality in sputtered SiO$_x$/Cu/SiO$_x$-based systems. By combining large-scale statistical analysis of more than 50,000 experimentally characterized devices with physics-based plasma and atomistic simulations, we show that device behavior does not emerge from deterministic process-to-performance mappings, but from a probabilistic cascade spanning defect formation, defect-state evolution, and functional-regime emergence. Data-driven clustering reveals a continuous functional state space composed of operational switching types, while inverse modeling identifies the reconstructed oxygen-vacancy density as an effective latent descriptor capturing the combined influence of structural disorder and defect topology. This latent descriptor is strongly coupled to both Cu redistribution and electrical response, linking otherwise hidden material properties to observable device characteristics. Furthermore, macroscopic switching behavior is argued to arise from ensemble integration across spatially heterogeneous subdomains, providing a physical explanation for the pronounced variability of large-area devices. These findings shift the perspective from deterministic defect engineering toward probabilistic defect-state design and establish a physically grounded framework for understanding and controlling functional variability in such oxide-based systems, such as memristive or resistive-switching devices.
\end{abstract}

\maketitle

\newpage


\section{Introduction}
\label{sec:introduction}

Resistive switching in oxide-based memristive devices has attracted extensive interest for emerging memory, in-memory computing, and neuromorphic applications \cite{Jin2025, Panda2024, Chaurasiya2023, Ielmini2016, Wong2012, Waser2007}. In these systems, device behavior arises from defect-mediated ionic and electronic transport processes occurring within structurally disordered oxide matrices, enabling highly nonlinear, history-dependent, and multifunctional electrical characteristics. At the same time, however, the nanoscale defect processes underlying these functionalities inherently introduce pronounced stochasticity and variability, which remain challenging to rationalize and predict from local defect formation to device-level behavior, particularly when prototyping novel multifunctional switching concepts \cite{Xiao2024, Song2023, Pan2014, Valov2013}.

The stochastic behavior of resistive oxide devices originates from the highly localized nature of the underlying switching processes. Their electrical response is typically governed by nanoscale ionic redistribution, defect generation, and the formation or rupture of conductive pathways within structurally heterogeneous oxide networks \cite{Ielmini2016, Valov2013}. As a result, small local variations in defect topology, stoichiometry, or transport pathways can induce large changes in macroscopic device behavior, including transitions between interface and filamentary type resistive switching regimes \cite{Atanasova2026, Pan2014}.

\begin{figure}
\includegraphics[width=8cm]{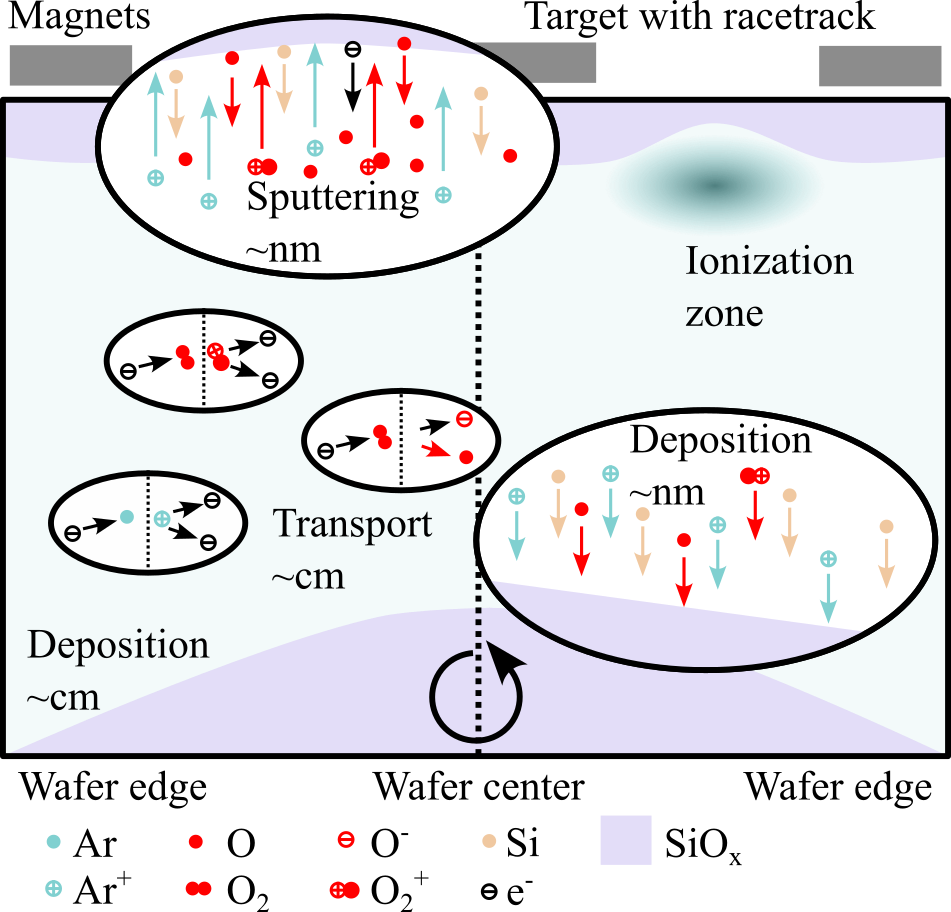}
\caption{Schematic illustration of the deposition process. The target and substrate processes are depicted at mesoscopic and nanoscale resolution, highlighting their governing role. In addition to elastic momentum transfer collisions, dominant transport processes are highlighted, i.e., Ar electron impact ionization, O\textsubscript{2} electron impact ionization, and O\textsubscript{2} dissociative attachment.}
\label{fig:deposition_schematic}
\end{figure}

Consequently, a central challenge is to understand how deposition conditions govern the formation of internal defect states and how these states ultimately determine device functionality \cite{Marquardt2022, Cipo2020}. As schematically illustrated in Figure~\ref{fig:deposition_schematic}, reactive thin-film growth processes simultaneously influence stoichiometry, defect generation, structural disorder, and local transport environments across multiple length scales \cite{Berg2005, Anders2010, Depla2024, Gudmundsson2022, Oehrlein2018}, while electrical switching behavior results from their coupled interaction during device operation \cite{Ielmini2016, Valov2013}. Hence, experimentally observed functionality often reflects the integrated outcome of complex and spatially heterogeneous internal material states rather than direct deterministic process-to-performance mappings. Integrated multiscale modeling and simulation approaches have increasingly connected reactor-scale plasma conditions with surface- and feature-scale processes (e.g., trench etching) \cite{gravesInfluenceModelingSimulation2003, mouchtourisMultiscaleModelingLow2017, denpohMultiscalePlasmaFeature2020}. However, extending such descriptions across the intermediate device scale (\textmu m--mm), while retaining the influence of atomistic-scale variations in material structure, remains an open challenge. Thus, although plasma growth, defect physics, ionic transport, and electrical switching mechanisms have each been studied extensively \cite{Gudmundsson2022, Cipo2020, Dittmann2021, Oehrlein2018, Ielmini2016, Valov2013}, the causal multiscale pathway linking plasma-defined growth conditions to atomistic defect formation and ultimately macroscopic device functionality remains largely unresolved \cite{Poehls2021, zahariCorrelationSputterDeposition2019, Marquardt2022}.

\begin{figure}
\includegraphics[width=8cm]{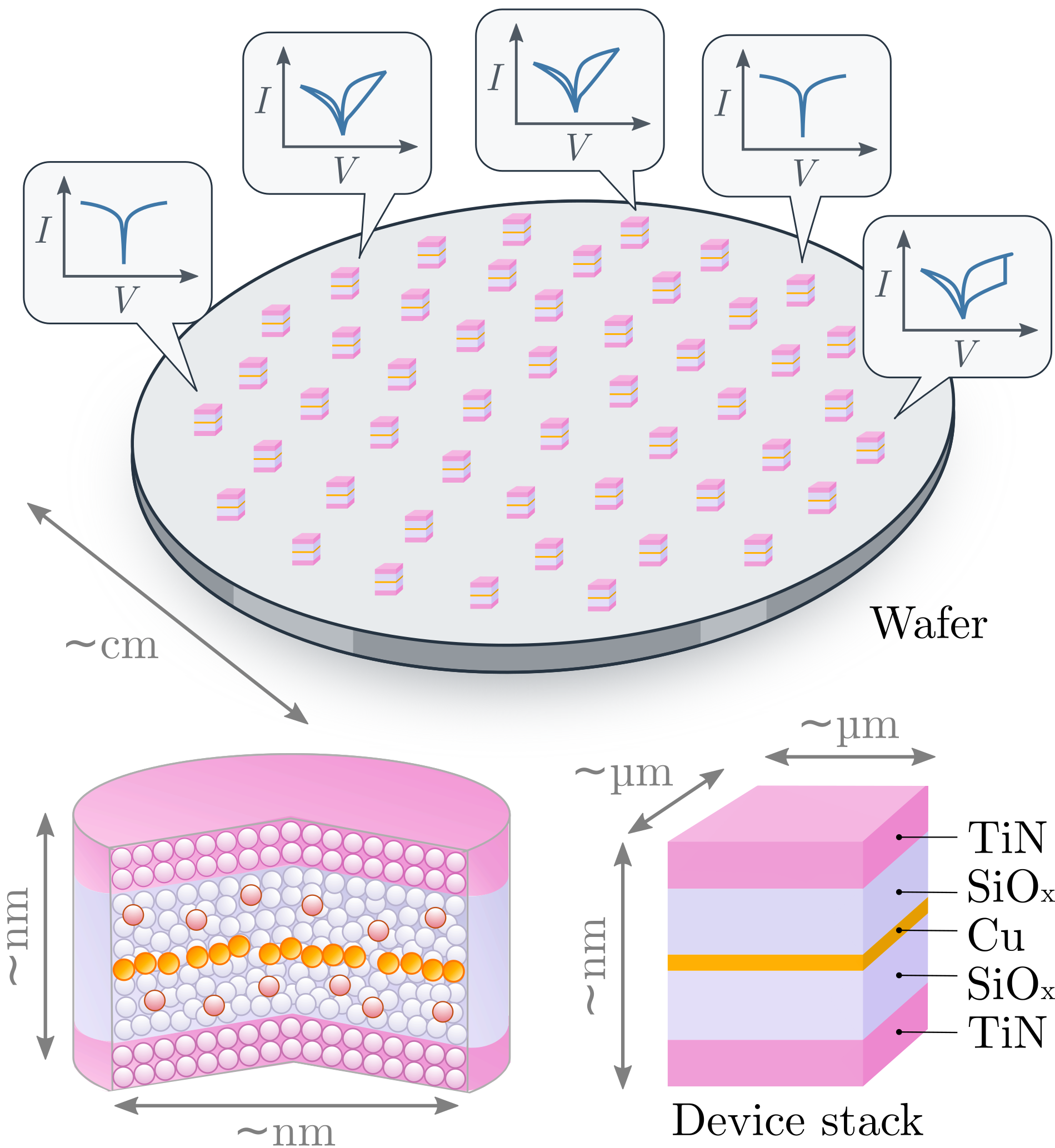}
\caption{Schematic illustration of wafer-scale device variability. Nominally identical devices distributed across the wafer can exhibit diverse electrical responses, represented by exemplary \mbox{\textit{I}--\textit{V}} characteristics. The TiN/SiO$_x$/Cu/SiO$_x$/TiN device architecture is shown at mesoscopic and nanoscale resolution, where red and yellow spheres indicates $V_\mathrm{O}$ and Cu, respectively.}
\label{fig:wafer_device_sketch}
\end{figure}

Among oxide-based (e.g., TiO$_x$~\cite{Islam2025}, HfO$_x$~\cite{Marquardt2023, Banerjee2022}, TaO$_{x}$~\cite{Prakash2013}, and SiO$_x$~\cite{Mehonic2018}) resistive switching systems, the TiN/SiO$_x$/Cu/SiO$_x$/TiN design stack, introduced most recently \cite{Lamprecht2025},  represents a particularly sensitive and, hence, interesting system for studying the interplay between defect-state formation, ionic redistribution, and electrical functionality. The device architecture and its associated electrical variability across the wafer are schematically illustrated in Figure~\ref{fig:wafer_device_sketch}. A sketch illustrating the device design and sensitivity (across scales) in shown in Figure~\ref{fig:wafer_device_sketch}. Previous studies on Cu-embedded SiO$_x$ memristive device demonstrated promising analog switching characteristics and revealed pronounced time- and temperature-dependent Cu redistribution and clustering phenomena (i.e., pancake-like Cu particles with diameters of the order of 100~nm), suggesting that internal Cu morphology evolution plays an important role in determining resulting device performance \cite{Lamprecht2026}. In parallel, device-level simulations suggested that oxygen-vacancy distributions and Cu incorporation can strongly influence the emergence of distinct electrical response characteristics across these systems \cite{Yarragolla2026}. Despite these individual findings, the holistic causal multiscale relationship between deposition conditions, atomistic defect formation, Cu redistribution after deposition, and macroscopic functionality is unclear both qualitatively and quantitatively.

\begin{figure}
\includegraphics[width=8cm]{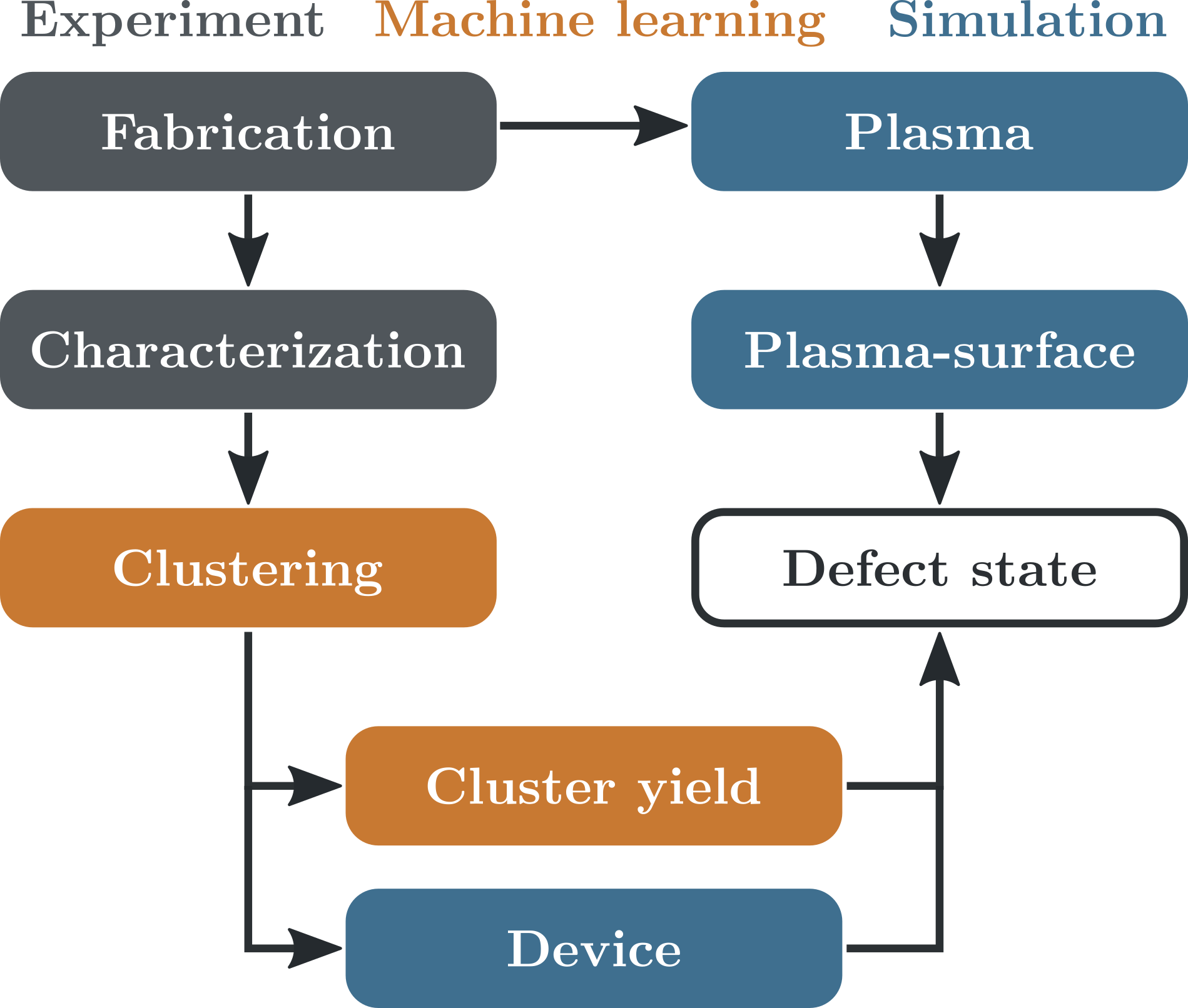}
\caption{Schematic illustration of the information flow underlying the forward and reverse engineering of defect-state formation in Cu-embedded SiO$_x$ memristive devices.}
\label{fig:info_flow}
\end{figure}

Here, we address this gap through complementary reverse- and forward-engineering pathways that interrogate different aspects of the underlying internal material state from opposite directions. The corresponding framework and information flow are summarized in Figure~\ref{fig:info_flow}. Reverse engineering combines experimental device characterization, data-driven analysis, and device-level simulations to reconstruct effective defect-state characteristics from macroscopic device behavior, whereas forward engineering employs plasma and atomistic simulations to resolve how plasma-defined growth conditions give rise to the corresponding material structure and defect states. By connecting these complementary descriptions, both pathways establish a multiscale link between processing conditions, atomistic material formation, and macroscopic device functionality.

The results presented in Section~\ref{sec:results} first establish the reverse-engineered relationship between device functionality and internal defect states and subsequently resolve the physical origins of these states through forward modeling. Both perspectives are integrated into a unified multiscale picture in the discussion of Section~\ref{sec:closure}. In Section~\ref{sec:conclusion} conclusions are drawn, whereas Section~\ref{sec:methods} presents the utilized methods in detail.


\section{Results}
\label{sec:results}

In this section, large-scale \textit{I}--\textit{V} device measurements are statistically analyzed to identify emergent electrical switching regimes. In combination with a reconstruction of the underlying defect-state landscape, these functional states are traced back to the origins of their physical processes. This establishes a cross-scale framework linking experimentally observed switching variability to latent defect states and their plasma- and atomistic-scale formation mechanisms.

\subsection{Emergent switching regimes and functional state space} 
\label{ssec:result_clusters}

\begin{figure*}
\includegraphics[width=16cm]{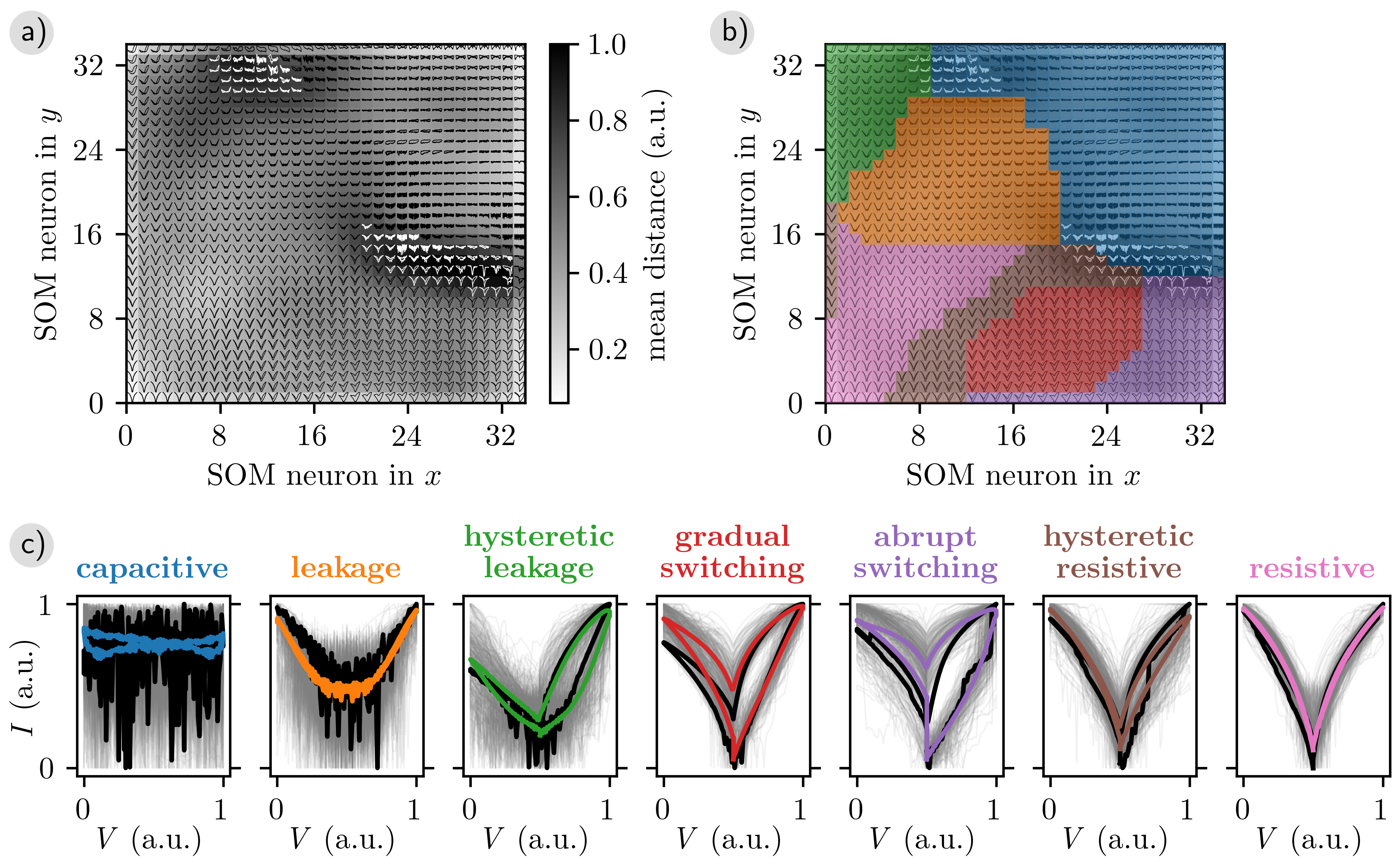}
\caption{a) U-matrix predicted by SOM, superimposed with the \textit{I}--\textit{V} curves centroids of the respective $34 \times 34$ clusters; b) Cluster aggregation based on \textit{I}--\textit{V} features; c) Per cluster: 200 randomly selected, experimental, and min-max normalized \textit{I}--\textit{V} curves (gray), superimposed with the results of the CIC device simulations (black) and with the centroids of all respective \textit{I}--\textit{V} curves (colored). An enlarged version of a) is included in the Supplementary Information for better readability. The color in b) and c) codes the key nanoelectrical characteristics.}
\label{fig:clusters_model}
\end{figure*}

To categorize the functional variation of sputtered SiO$_x$ devices, 50,000 individual \textit{I}--\textit{V} characteristics were analyzed using self-organizing maps (SOMs), yielding 1,156 representative variability clusters that capture the intrinsic organization of the switching landscape. The resulting U-matrix (inter-neuron distance map; Figure~\ref{fig:clusters_model}~a) reveals a highly structured functional state space, in which contiguous low distance regions (bright) correspond to related switching responses, while pronounced high distance boundaries (dark) separate functionally distinct behavioral domains.

Hierarchical aggregation of this state space further condenses the observed variability into seven emergent switching regimes (Figure~\ref{fig:clusters_model}~b), each defined by distinct electrical response classes. Representative regime centroids (colored) are compared in Figure~\ref{fig:clusters_model}~c) to experimental data (gray) and physics-based cloud-in-a-cell (CIC) simulations (black), with mean percentage deviations of 16.5~\% (experiment-centroid) and 17.1~\% (simulation-centroid), supporting both the statistical robustness and physical interpretability of the derived regime taxonomy.

With the exception of one regime, denoted ``hysteretic resistive'' and associated with asymmetric hysteresis during either the set or reset phase, the aggregated regimes occupy largely contiguous domains within the SOM topology. Transition zones between neighboring regimes emerge systematically (cf. Figure~\ref{fig:clusters_model}~a) and exhibit mixed electrical characteristics, indicating that device variability is governed not by strictly discrete categories, but by a continuous stochastic functional landscape with manifold-like organization around a finite set of emergent switching regimes. Notably, capacitive-like switching remains strongly segregated by pronounced topological boundaries (i.e., high mean distance in Figure~\ref{fig:clusters_model}~a), identifying it as a fundamentally distinct operational regime, with leakage behavior constituting its primary transitional interface.

\subsection{Physics-informed inference of latent defect state landscapes}
\label{ssec:result_process_device}

The emergent switching regimes identified in Section~\ref{ssec:result_clusters} suggest that the observed device variability is governed by an underlying low dimensional latent defect state structure. To elucidate this relationship, a physics-informed inverse modeling approach is employed to reconstruct the internal defect state space from the process information underlying the experimental \textit{I}--\textit{V} characteristics and device-level simulations.

\begin{figure*}
\includegraphics[width=16cm]{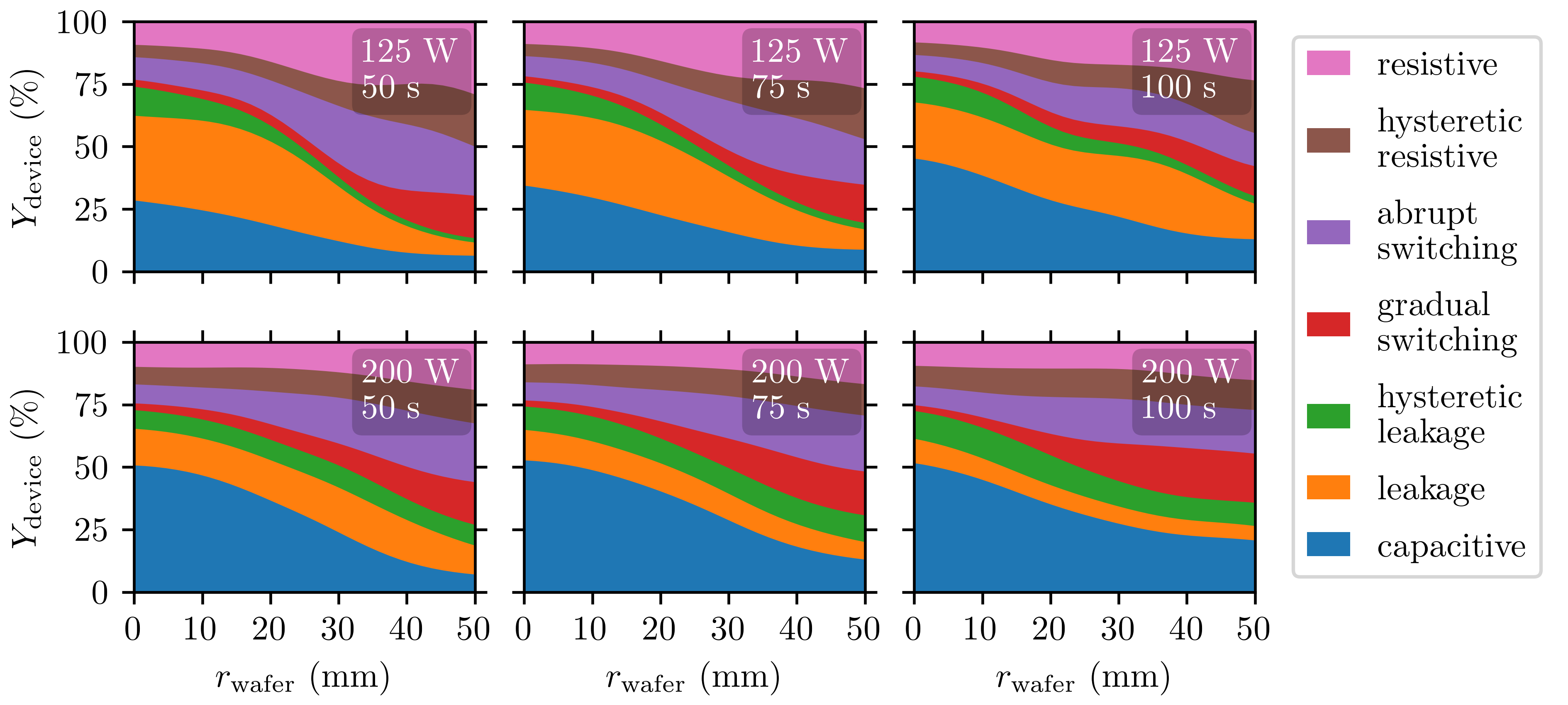}
\caption{Generalized manufacture yields $Y_\mathrm{device}$ as a function of the radial device site on the wafer $r_\mathrm{wafer}$ during the SiO$_x$ deposition. Power $P_{\mathrm{SiO}_x}$ and deposition time $t_{\mathrm{SiO}_x}$ are included in each top right corner.}
\label{fig:KDE_results}
\end{figure*}

The high dimensional process parameter space is reduced via a sensitivity analysis (cf.~Section~\ref{sssec:method_ML_cluster_yield}), which identifies the three most influential variables governing device variability: radial device position on the wafer $r_\mathrm{wafer}$, sputtering power $P_{\mathrm{SiO}_x}$ and deposition time $t_{\mathrm{SiO}_x}$.

As shown in Figure~\ref{fig:KDE_results}, cluster yield distributions reveal a pronounced dependence on wafer position and deposition conditions. Notably, substantial device variability persists both within and across wafers at identical nominal process settings, as further illustrated by representative spatial cluster distributions in the Supporting Information. The pronounced within-wafer variability indicates a substantial stochastic contribution to the switching behavior.

Deposition time and power provide secondary modulation of the regime populations. Resistive device fractions decrease systematically with increasing deposition time across all conditions. In contrast, memristive (i.e., hysteretic-based, switching-based) and capacitive regimes exhibit complementary, power dependent behavior: increasing deposition time shifts device populations toward capacitive regimes at 125 W, but toward memristive regimes at 200 W, revealing a power induced inversion of dominant switching pathways.

\begin{figure}
\includegraphics[width=8cm]{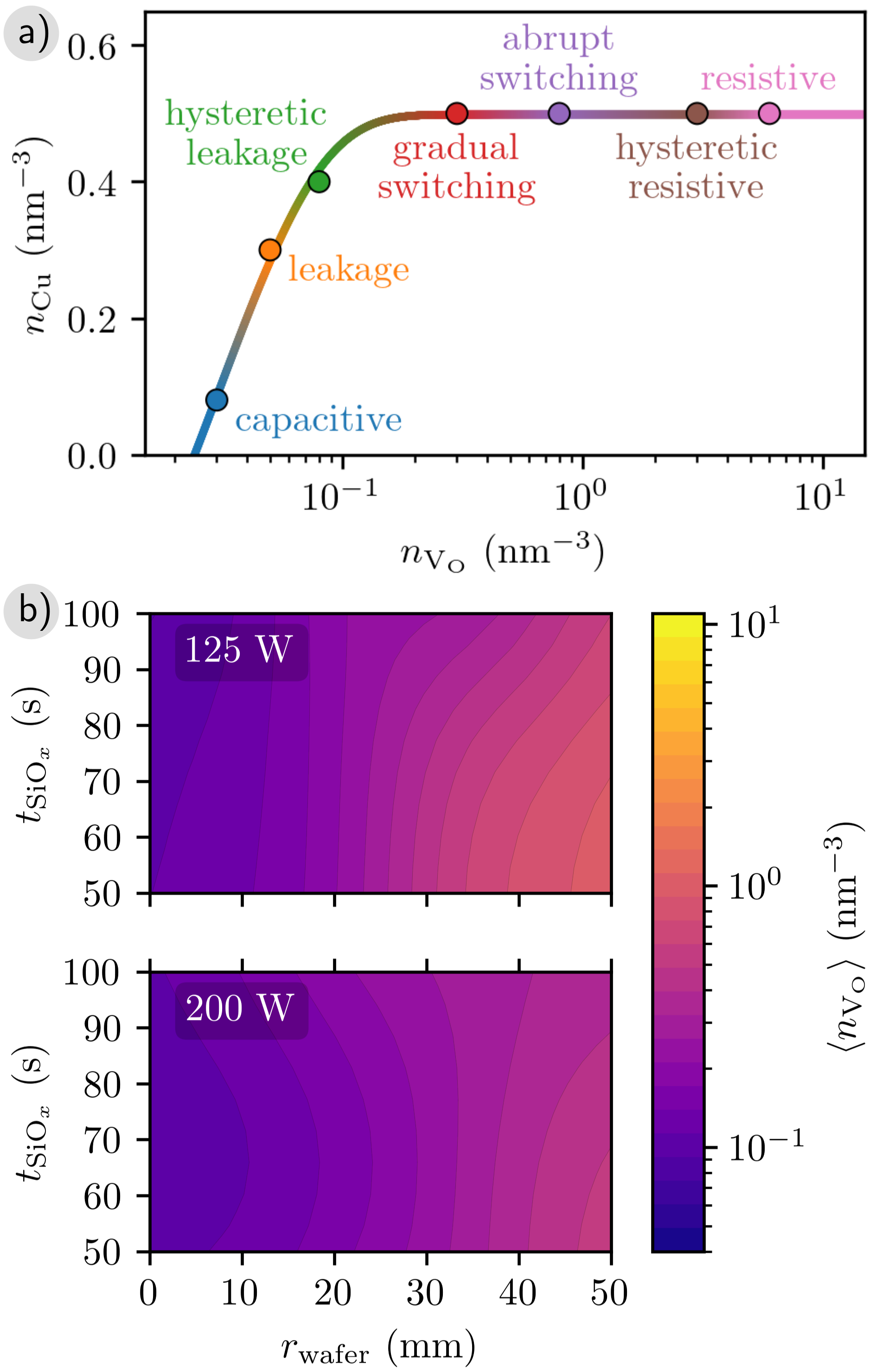}
\caption{a) In SiO$_x$ dissolved effective Cu density $n_\mathrm{Cu}$ as a function of the effective oxygen vacancy density $n_\mathrm{V_O}$ required by CIC device-level simulations. b) The expected value of the oxygen vacancy densities as a function of the radial device site on the wafer $r_\mathrm{wafer}$ and deposition time $t_{\mathrm{SiO}_x}$. The power $P_{\mathrm{SiO}_x}$ is included in each top left corner.}
\label{fig:device_state}
\end{figure}

Together, these observations indicate that process conditions do not deterministically define individual switching states, but instead govern the statistical occupation of distinct device state populations, reflecting an underlying process-dependent evolution of the defect state landscape. To reconstruct this statistical mapping, each switching regime is assigned a corresponding effective internal defect-state representation retrieved from device-level simulations by reconstructing corresponding $I-V$ characteristics. Figure~\ref{fig:device_state}~a) reveals a correlation between Cu density $n_\mathrm{Cu}$ and oxygen vacancy density $n_\mathrm{V_O}$ across the seven regimes, indicating that both quantities are not independent descriptors but instead arise from a shared underlying defect state trajectory. This relationship is well approximated by a saturated exponential dependence (cf. line plot in Figure~\ref{fig:device_state}),
\begin{equation}
n_\mathrm{Cu} = n_\mathrm{Cu}^\mathrm{max} \left[ 1-\exp(-k(n_\mathrm{V_O} - n_\mathrm{V_O}^\mathrm{min})) \right],
\end{equation}
enabling a reduction of the effective latent state dimensionality. With a relative error of 2.53~\%, the coefficients were determined to equal $n_\mathrm{Cu}^\mathrm{max}=0.50$~nm$^{-3}$, $k=32.7$~nm$^3$ and $n_\mathrm{V_O}^\mathrm{min}=0.02$~nm$^{-3}$. Based on this reduced representation, oxygen vacancy density is used as the principal defect state coordinate in the following analysis. By weighting the regime-specific defect states (densities) with the experimentally observed cluster yield distributions, an ensemble averaged defect state expectation value is obtained for each process condition. The resulting defect state landscapes, shown in Figure~\ref{fig:device_state}~b), reveal pronounced radial gradients and power dependent evolution, providing a physically interpretable bridge between process space and functional device variability.

The reconstructed ensemble averaged oxygen vacancy landscapes reveal wafer position as the dominant determinant of defect state evolution across all investigated process conditions. In both power regimes, defect densities remain comparatively low and of similar magnitude within the wafer center (approximately $r_\mathrm{wafer} \lesssim 20$–25 mm), indicating a robust low defect process window largely independent of deposition conditions. Beyond this region, oxygen vacancy densities increase nonlinearly toward the wafer edge, defining a pronounced radial defect amplification. 

This radial defect-density trend is consistent with the observed distribution of switching types and agrees with previous studies on memristive devices, where the oxygen-vacancy density is known to strongly influence the transition between regimes \cite{Wu2026, Gul2017}. Low reconstructed vacancy densities, particularly in the wafer-center region and more pronounced for the 200~W condition, correlate with an increased occurrence of capacitive-like and leakage-like functionalities (cf.~Figure~\ref{fig:KDE_results}), whereas the stronger edge-directed defect amplification at 125~W is associated with more resistive behavior. The predominance of memristive-family curves near the 200~W wafer edge therefore suggests an intermediate defect-density window that is favorable for memristive switching.

This edge directed amplification is strongly power dependent. At lower power, that is 125 W, the radial defect increase emerges earlier (approximately 20~mm), exhibits a steeper nonlinear rise, and reaches substantially higher edge defect densities than at 200~W, where comparable amplification is shifted to larger radii (approximately 35~mm) and remains significantly attenuated. Increasing deposition time primarily acts as a secondary modulation mechanism, exerting limited influence in central wafer regions but progressively suppressing defect accumulation at the wafer's edge, particularly under low power conditions.

Collectively, these results identify power as a regulator of the spatial onset and severity of radial defect amplification, while deposition time predominantly mitigates edge localized defect accumulation. This emphasizes that wafer position is the primary axis of defect state variability, with process parameters governing the expansion, suppression, and radial displacement of high defect regimes rather than uniformly shifting the overall defect landscape.

\subsection{Plasma and atomistic origins of defect formation}
\label{ssec:Plasma_and_atomistic}

To identify the physical mechanisms underlying the reconstructed defect state landscapes, plasma simulations provide an estimate of the spatial distribution of growth conditions across the wafer, including local stoichiometry, ion flux, and incident energy distribution at the surface.

\begin{figure}
\includegraphics[width=8cm]{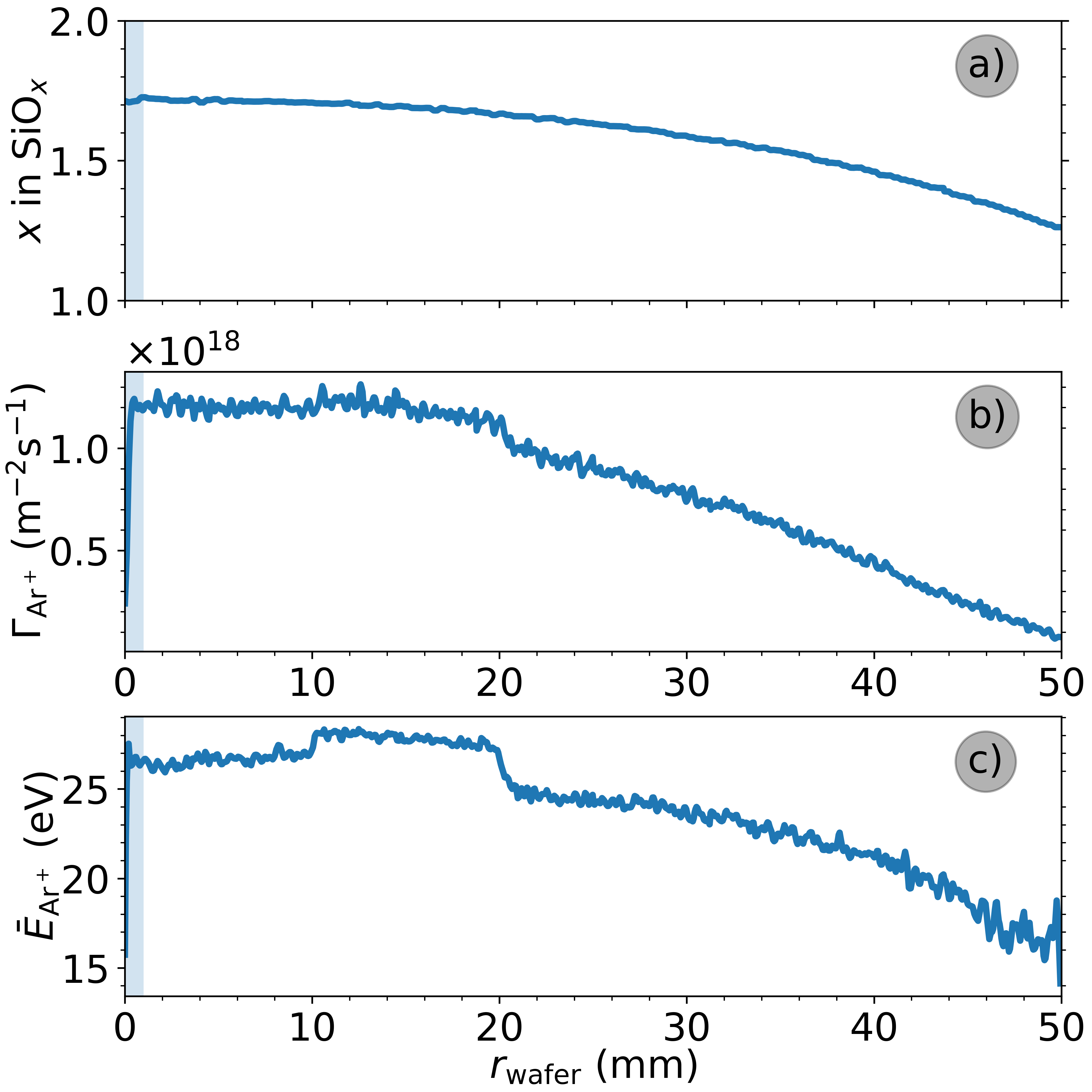}
\caption{Results of Ar/O$_2$ plasma simulations at 5 \% O$_2$, 0.5 Pa, 300 K, $V_\mathrm{rf}=250$ V, and $V_\mathrm{SB}=-110$ V. a) Chemical composition $x$ in SiO$_x$, b) Ar$^+$ ion flux $\Gamma_\mathrm{Ar^+}$ , and c) average energy per Ar$^+$ ion $\bar{E}_\mathrm{Ar^+}$ at the substrate as a function of radial position.}
\label{fig:plasma_simulation_1}
\end{figure}

As shown in Figure~\ref{fig:plasma_simulation_1}~a), these simulations reveal a pronounced radial non-uniformity across the substrate surface. The central wafer region (up to 20~mm radius) exhibits approximately constant SiO$_x$ stoichiometry of 1.7, beyond which oxygen content decreases progressively toward the wafer edge to 1.25. This compositional gradient is primarily governed by the competing mechanism of sputtered Si deposition and oxidation from impinging O atoms (due to O$_2$ electron impact dissociation) and O$_2^+$ ions from the plasma bulk. Assuming a stepwise oxidation of the surface sites s-Si and s-SiO during oxygen impingement and reduction of s-SiO and s-SiO$_2$ during Si impingement, the effective stoichiometry is locally balanced. At the opposing target surface, the oxygen content varies with the lowest surface oxidation state in the racetrack region, as expected, with values ranging between 0.9 at the center of the racetrack up to 1.8 at the target center (target poisoning, see Supplementary Information).

Figure~\ref{fig:plasma_simulation_1}~b) depicts the Ar$^+$ ion flux, which remains approximately constant near the wafer center (up to 20~mm radius) and from thereon decreases linearly to zero toward the outer edge. The corresponding Ar$^+$ ion kinetic energy at the substrate surface (Figure~\ref{fig:plasma_simulation_1}~c) remains rather constant in the central region with $25-27$~eV per Ar$^+$ ion, but decreases steadily toward the wafer edge, where Ar$^+$ ions arrive with kinetic energies of up to 17~eV. A small increase is observed between $r=10$ and 20~mm, reflecting the spatial distribution of the electric potential in the magnetized plasma region where negative oxygen ions O$^-$ are confined and argon ions Ar$^+$ are created (see Supplementary Information). In addition, the energy per impinging electron is observed in the range from 10~eV to 26~eV, which may further contribute to local structural evolution, with thermalization timescales ranging from 1 ps in oxygen-rich SiO$_x$ to 10 ps in Si-rich regions (see Supplementary Information). These electron processes are unlikely to constitute the primary origin of the observed structural modifications, although they may contribute to short-lived local relaxation processes. However, electron injection can facilitate aggregation of oxygen vacancies \cite{Gao2016, Gao2019}, whose relevance has been discussed in the preceding section.

Together, these spatially varying compositional and energetic growth conditions define the substrate specific ion bombardment environment for atomistic defect formation. Notably, the plasma simulations were found to underestimate the absolute SiO$_x$ deposition by approximately one order of magnitude with respect to experimental deposition rates. This discrepancy is primarily attributed to an underestimated sputtering yield which is strongly nonlinear in the low energy regime around 100~eV (of the order of the average self-bias voltage $V_\mathrm{SB}=-110$~V). Additionally, this may relate to modeling uncertainties along the target-to-substrate pathway, including target sputtering, gas-phase processes, and net incorporation at the growing film. In contrast, the simulated Ar$^+$ ion flux and the SiO$_x$ stoichiometry represent more direct or relative quantities and are therefore expected to be substantially less sensitive to such cumulative uncertainty propagation. Hence, $\Gamma_{\mathrm{Si}}$ was derived from experimentally measured growth rates, placing the experimentally relevant Ar$^+$-to-Si flux ratio at approximately unity within the parameter space explored by the atomistic plasma-surface simulations described below.

\begin{figure*}
\includegraphics[width=16cm]{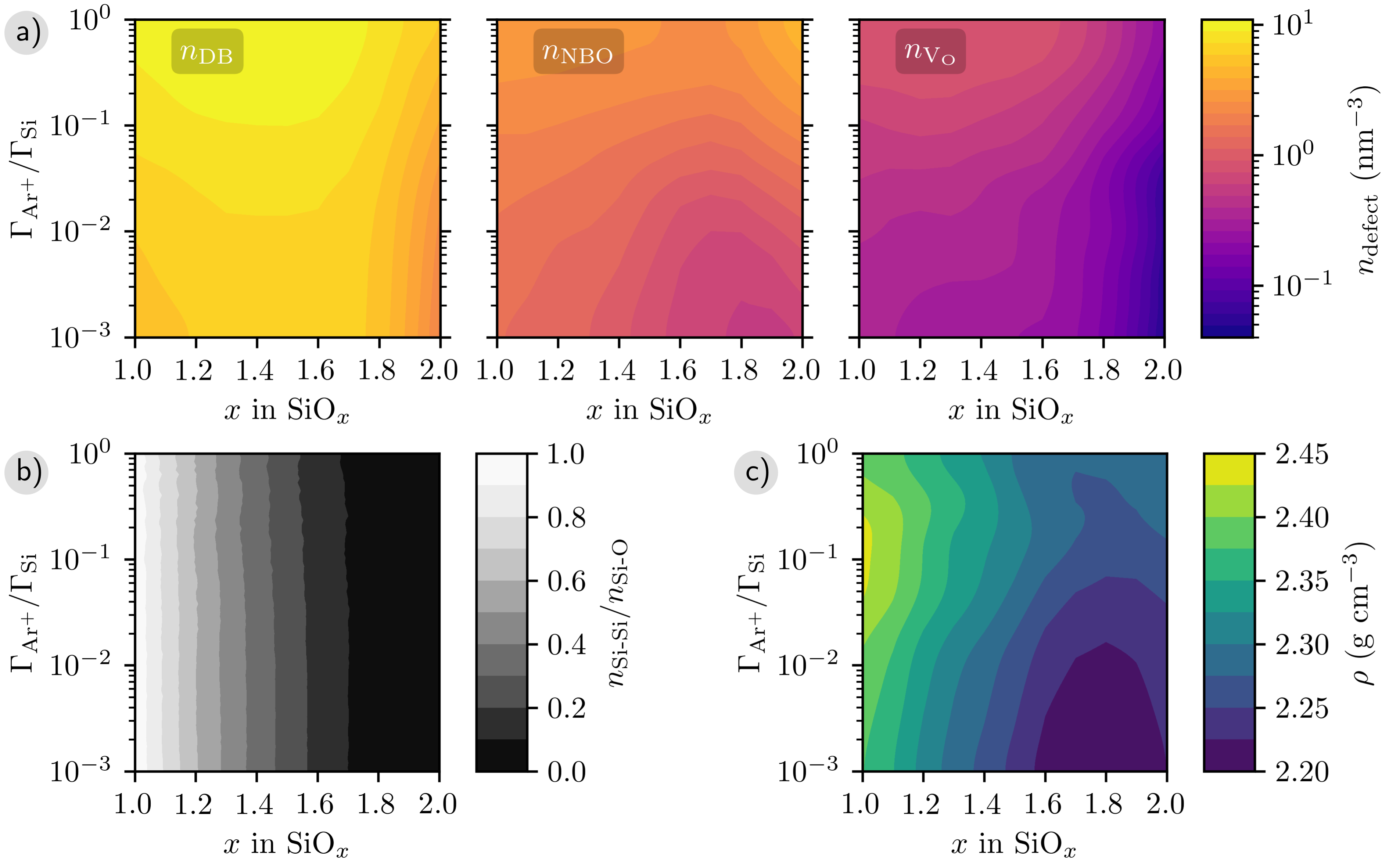}
\caption{Results of Ar ion-SiO$_x$ interaction simulations for $x \in [0,1]$ in SiO$_x$ and Ar$^+$ flux to Si flux ratio~$\Gamma_\mathrm{Ar^+}/\Gamma_{\mathrm{Si}} \in [0,1]$ with mean ion energies of 35~eV. a) Intrinsic defect densities for DBs, NBOs, and V$_\mathrm{O}$. b) Ratio of Si-Si bond density to Si-O bond density~$n_\text{Si-Si}/n_\text{Si-O}$. c) Mass density~$\rho$.}
\label{fig:MD_results}
\end{figure*}

To identify the atomistic origin of ion bombardment induced defect formation, molecular dynamics (MD) simulations of Ar$^+$ interactions with SiO$_x$ ($1 \leq x \leq 2$) were performed. The simulations capture the interplay between ion energy deposition, local stoichiometry, and defect evolution across the amorphous network. Across all stoichiometries, the average incorporated Ar concentration remains nearly constant at 1.23 $\pm$ 0.37 nm$^{-3}$, indicating that the observed structural transitions are governed primarily by the network specific response to ion bombardment rather than differences in Ar uptake itself.

Figure~\ref{fig:MD_results}~a) reveals that silicon dangling bonds (DBs), non-bridging oxygen defects (NBOs), and oxygen vacancies V$_\mathrm{O}$ occupy systematically distinct density scales, approximately separated by one order of magnitude each. In particular, the comparable range of oxygen vacancy densities obtained from MD and effectively derived from the device simulation ensembles in the previous Section~\ref{ssec:result_process_device} provides qualitative cross-scale consistency, despite the fundamentally different physical descriptions of the two approaches. However, the defects' relevance is not determined by abundance alone: while NBOs and Si-Si bond formation predominantly govern structural densification and mechanical response, V$_\mathrm{O}$ remain critical for memristive defect functionality (cf. \ref{ssec:result_process_device}).

As shown in Figure~\ref{fig:MD_results}~b), the Si-Si to Si-O bond ratio identifies a stoichiometry driven topological transition near $x \approx 1.7$, separating two distinct atomistic defect formation regimes.

For O-rich compositions ($1.7 \leq x \leq 2.0$), the amorphous SiO$_2$-like network remains structurally intact and largely free of Si-Si bonds \cite{Ugwumadu2023}. In this regime, decreasing oxygen content primarily introduces oxygen vacancies, which partially relax into dangling bonds while preserving the underlying tetrahedral network topology.

Ion bombardment therefore acts predominantly through Si-O bond disruption, driving network densification, compressive stress accumulation, and increasing NBO formation.

For SiO$_2$-like compositions, the simulated mass density (Figure~\ref{fig:MD_results}~c) and residual stress (between $-53.48$~MPa and $-456.45$~MPa; Supplementary Information) remain within experimentally reported ranges for ion-assisted sputtered silica, including densities of 1.91--2.38~g\,cm$^{-3}$ and residual stresses between $-1$~MPa and $-440$~MPa \cite{Simurka2018}, as well as reported stress values of $-90$~MPa to $-300$~MPa under related sputter-deposition conditions \cite{Bhatt2007}, supporting the physical plausibility of the reconstructed response mode.

Below $x \approx 1.7$, progressive Si-Si bond formation marks a transition toward a Si-rich regime in which oxygen deficiency is accommodated through network reconstruction rather than isolated defect formation alone.

This transition fundamentally alters the material response: mass density becomes increasingly governed by composition itself, consistent with experimentally observed reductions in density and increasing open volume with increasing oxygen deficiency in SiO$_x$ \cite{Brusa2003}, while intrinsic defect populations emerge even at minimal ion doses. Such composition-driven defect formation is consistent with the increasingly structure-dependent nature of oxygen-deficient configurations in amorphous oxides, where local coordination and network topology can intrinsically accommodate oxygen deficiency \cite{Strand2024}. Ion bombardment therefore acts on an already defect-enriched and structurally modified network, resulting in more immediate and compositionally offset defect accumulation.

This stoichiometry driven transition distinguishes a densification dominated silica-like regime from a topology reconstruction dominated Si-rich regime, demonstrating that ion bombardment interacts not with a single amorphous material class, but with fundamentally different atomistic response pathways depending on network composition. These distinct formation mechanisms provide the atomistic basis for the previously reconstructed defect state landscapes and their process-dependent functional variability.

\section{Discussion}
\label{sec:closure}

The preceding multiscale analyses establish that the Cu-embedded SiO$_x$ memristive device functionality is not governed by a direct deterministic process-to-performance mapping, but instead emerges through a probabilistic cross-scale cascade linking plasma-defined growth conditions, atomistic defect formation, latent defect-state evolution, and stochastic functional regime occupation. Rather than prescribing singular electrical states, process parameters reshape the statistical accessibility of specific material response pathways, thereby biasing the likelihood of distinct internal device states. While such functional variability is widely associated with nanoscale defect heterogeneity, structural inhomogeneity, and probabilistic conductive pathway formation in resistive oxide systems \cite{Waser2007,Ielmini2016,Wong2012,Valov2013,Pan2014}, the present framework links this variability to plasma-defined growth conditions and the resulting hierarchy of atomistic and latent defect-state formation.

At the process level, wafer-scale plasma non-uniformity defines the dominant axis of device variability by spatially modulating local a-SiO$_x$ stoichiometry and the ion-to-deposited-particle ratio across the substrate. Plasma simulations reveal a comparatively uniform central growth regime around a-SiO$_{1.7}$, followed by a progressive shift toward increasingly oxygen-deficient edge conditions ($x\approx1.25$), consistent with gradients generally observed in reactive sputtering processes \cite{Berg2005, deplaReactiveSputterDeposition2008, zahariCorrelationSputterDeposition2019, Marquardt2022}. MD further demonstrates that this plasma-defined compositional evolution crosses a stoichiometry-dependent transition between distinct atomistic response regimes near $x\approx1.7$, close to previously reported composition-driven transitions in the electronic and network characteristics of amorphous SiO$_x$ near $x\approx1.5$ \cite{Bell1988}. Specifically, the response changes from a densification-dominated silica-like regime toward a structurally softened Si-rich network characterized by topology reconstruction, enhanced defect accessibility, and altered mechanical response. The silica-like response is consistent with previous atomistic simulations of ion-assisted SiO$_2$ growth at similar impact energies, where densification proceeded predominantly through intermediate-scale network-topology modifications while largely preserving local tetrahedral building blocks \cite{Lefevre2001}. Within this framework, power regulates where and how strongly the defect-state transition develops, with lower power enabling earlier and more pronounced radial defect amplification, whereas higher power shifts its onset outward and suppresses its severity.

These growth-induced atomistic regimes subsequently define the effective defect-state landscape governing copper redistribution within the SiO$_x$ matrix. Importantly, the reconstructed oxygen-vacancy coordinate should not be interpreted as an isolated vacancy density alone \cite{Strand2024}, but rather as an effective latent descriptor that compresses the combined influence of intrinsic defects, structural disorder, and network compliance into a reduced functional state variable. The observed saturated exponential coupling between this effective defect state and copper density suggests a defect-enabled but self-limiting redistribution process, in which increasing defect accessibility promotes copper incorporation and internal restructuring until finite uptake, transport, or structural accommodation limits impose saturation. This interpretation is consistent with the strong dependence of copper transport and clustering on the local structure and trapping environment of SiO$_2$-based matrices \cite{Guzman2015}. Moreover, the diffusion-driven formation of extended copper-rich “pancake”  structures observed in the same SiO$_x$/Cu/SiO$_x$ material system provides a direct morphological manifestation of such redistribution \cite{Lamprecht2026}. Together, these observations link growth-induced defect topology to copper redistribution and, ultimately, the formation of electrically active internal states, although the underlying atomistic transport mechanisms remain to be resolved.

Macroscopic switching behavior ultimately emerges from spatial integration across heterogeneous local defect states, transforming continuous nanoscale variability into structured but probabilistic functional device classes. The continuous transitions observed throughout SOM space indicate an underlying continuum of accessible states, while the seven identified switching regimes represent operational families within this broader landscape, consistent with the stochastic nature of resistive switching in oxide systems \cite{Ielmini2016,Wong2012}. Given that the lateral dimensions of the investigated devices exceed plausible defect-correlation lengths by several orders of magnitude, their electrical characteristics are therefore better understood as ensemble-integrated responses of heterogeneous local subdomains rather than homogeneous single-state systems. Within this framework, shifts in the statistical occupation of local states give rise to transitions between capacitive, memristive, and resistive responses, consistent with device-level modeling showing that the interplay between deep oxygen-vacancy-related traps and shallower copper trap states governs the resulting electrical behavior \cite{Yarragolla2026}. Localized high-defect regions may further bias global device behavior from interface-based toward filamentary operation when conductive pathways become dominant, consistent with the highly localized nature of conductive filaments in resistive switching devices \cite{Celano2014}. This reconciles the continuous functional manifold with discrete regime classes and provides a physical basis for the pronounced variability among nominally similar devices.

From an engineering perspective, these findings imply that functional control in sputtered SiO$_x$/Cu/SiO$_x$-based systems is best achieved not by targeting individual defect species, but by tuning the distribution of accessible defect-topology configurations across scales. Wafer position primarily selects the local growth regime through variations in SiO$_x$ stoichiometry and the ion-to-deposited-particle ratio, power shifts the onset and severity of defect-state transitions, and deposition time modulates the statistical occupation within these regimes. Large-area devices therefore naturally exhibit probabilistic multifunctionality, whereas scaling toward smaller active volumes is expected to reduce ensemble averaging and promote more deterministic functional specialization. The weak dependence on device area within the investigated 100--2500~µm$^2$ range suggests that even the smallest studied geometries remain above characteristic defect-correlation scales \cite{Lamprecht2026}. More broadly, deposition strategies, such as atomic layer deposition,  offering tighter control over local material formation may therefore provide a route toward reduced stochasticity and more precise functional-regime engineering.


\section{Conclusion}
\label{sec:conclusion}

This work establishes a multiscale framework linking plasma-defined growth conditions, atomistic defect formation, and macroscopic functionality in Cu-embedded SiO$_x$ memristive device. By combining complementary reverse- and forward-engineering approaches, we show that device behavior does not follow a direct deterministic process-to-performance mapping, but emerges through a probabilistic cascade of material-state formation and functional regime occupation. Central to this framework is an effective latent defect-state description that connects growth-induced structural and defect heterogeneity with copper redistribution and the resulting electrical response.

Across these scales, wafer position emerges as the dominant source of systematic variability through plasma-induced gradients in SiO$_x$ stoichiometry and the ion-to-deposited-particle ratio, while power and deposition time further modulate the resulting defect-state distributions. Atomistic simulations reveal a stoichiometry-dependent transition between distinct material-response regimes, providing a physical origin for the reconstructed defect-state landscape. At the device scale, spatial integration across heterogeneous local states translates this continuous material variability into probabilistic occupation of distinct functional regimes.

More broadly, these findings shift the perspective from deterministic control of individual defect species toward probabilistic engineering of accessible material states. This provides a unified basis for understanding how process-induced variability propagates across scales into distinct device functionalities and establishes a route toward more controlled functional-state engineering in oxide-based memristive systems.


\section{Methods}
\label{sec:methods}

To investigate the multiscale relationship between plasma-defined growth conditions, defect-state formation, and macroscopic device functionality in sputtered SiO$_x$/Cu/SiO$_x$ systems, a combined experimental, data-driven, and physics-based modeling approach was employed. As schematically illustrated in Figure~\ref{fig:info_flow}, (heterogeneous) experimental device data obtained over multiple years of empirical process optimization were statistically organized through clustering and functional-state mapping (cluster yields), enabling probabilistic reverse engineering of internal defect-state characteristics using device-level simulations. Complementarily, forward modeling of defect-state formation was performed using plasma and plasma-surface interaction simulations, on the macroscopic and atomistic scale, respectively, to establish physically interpretable links between deposition conditions, local defect topology formation, and resulting material response. Together, forward and reverse engineering approaches establish a common defect-state representation, linking experimentally observed device behavior with growth-induced defect formation across scales.

\subsection{Device fabrication}
\label{ssec:methods_fabrication}

A statistically heterogeneous experimental device space comprising more than 50,000 devices designed as TiN/SiO$_x$/Cu/SiO$_x$/TiN stack and fabricated across 50 individual 4-inch wafers was investigated in this work. The dataset was accumulated over three years of empirical process optimization and spans broad variations in deposition conditions, device geometries, and post-fabrication degradation times.

In the following, the key aspects of device fabrication relevant to the present study are outlined. Further details on the device architecture, manufacture, and systematic characterization are provided in Ref.~\cite{Lamprecht2026}.

Device fabrication followed a CMOS-compatible process flow. Thermally oxidized 4-inch Si wafers were structured by lithography and etching into electrically isolated device regions with lateral areas $A_\mathrm{device}$ ranging from 100--2500~$\mu\textrm{m}^2$. Within these regions, TiN/SiO$_x$/Cu/SiO$_x$/TiN stacks were deposited by sequential RF~magnetron sputtering operated at 13.56~MHz using a target-to-substrate distance of 7~cm.

The SiO$_x$ layers were sputtered from a pure Si target at powers $P_{\mathrm{SiO}_x}$ of 100--200~W in reactive Ar/O$_2$ plasmas with oxygen flow fractions $F_{\mathrm{O}_2}/F_\mathrm{total}$ ranging from 5--25~\% and deposition times $t_{\mathrm{SiO}_x}$ between 32--145~s (to explore the influence of different SiO$_x$ thicknesses). Without breaking vacuum, a sub-monolayer Cu interlayer was subsequently sputtered from a Cu target operated at 40~W in pure Ar plasma for 1.3--1.9~s. Processing was completed by SiO passivation and TiN metallization (for electrical contacting) using thermal evaporation and Ar/N\textsubscript{2} reactive direct-current (DC) sputtering, respectively.

\subsection{Reverse engineering}
\label{ssec:method_reverse}

The reverse-engineering workflow aims to reconstruct effective internal defect-state characteristics from experimentally observed device behavior. For this purpose, electrical measurements, statistical clustering, probabilistic state mapping, and device-level simulations were combined to establish a common functional representation, linking macroscopic electrical response to latent defect-state evolution.

\subsubsection*{Electrical characterization}
\label{ssec:methods_measurements}

Electrical characterization was performed on wafer level using a semi-automated probe station equipped with an HP~4156A semiconductor parameter analyzer. For each measurement cycle, bipolar voltage sweeps were applied. Measurement times per voltage step, $t_\mathrm{measure}$, ranged from 0.1--1~s, while maximum voltage amplitudes $U_\mathrm{sweep}$ varied between 1--1.8~V depending on device behavior and measurement protocol. To capture both immediate and time-dependent functional variability, post-fabrication degradation times $t_{\mathrm{deg}}$ between 3~days and 3~months were included within the investigated experimental parameter space. Additional electrical characterization details are provided in Ref. \cite{Lamprecht2026}.

\subsubsection*{Clustering of $I$--$V$ characteristics}
\label{sssec:method_ML_clustering}

A total of more than 50,000 experimentally measured $I$--$V$ characteristics obtained from 50 wafers were statistically evaluated to establish a reduced functional-state representation of the heterogeneous experimental device space. Depending on the applied protocol, individual measurements contained up to 203 data points. Curves containing fewer points (e.g., 103) were aligned by subsampling (i.e., piecewise linear interpolation) to a common representation.

Substantial device-to-device variability was observed across the dataset, requiring abstraction of the individual $I$--$V$ curves into a lower-dimensional functional representation. Multiple clustering approaches were considered but failed to produce physically interpretable and functionally coherent cluster representations.

Specifically, k-means \cite{MacQueen1967}, spectral clustering \cite{Ng2001}, hierarchical density-based spatial clustering of applications with noise (HDBSCAN) \cite{Campello2013}, all implemented in scikit-learn \cite{Pedregosa2011}, were evaluated with and without prior dimensionality reduction (e.g., principal component analysis \cite{Pearson1901}) and using different feature scaling strategies (e.g., min-max normalization, Z-score standardization) in both original and log-transformed feature spaces. Ultimately, the self-organizing map (SOM) method was adopted \cite{Kohonen1982}. This approach was selected due to its simultaneous dimensionality reduction and topology-preserving clustering capabilities.

Prior to clustering, the data were preprocessed in three steps. First, the voltage input was normalized by the applied sweep range $U_\mathrm{sweep}$, resulting in a normalized voltage coordinate spanning $-1$ to $1$ for all measurements. The current response was subsequently processed independently. To avoid singularities during logarithmic transformation, the current region corresponding to input voltages between $-0.1$~V and $0.1$~V was excluded. The logarithm of the current magnitude was then calculated to account for the strong skewness of the data spanning current levels between approximately $10^{-10}$~A and $10^{-1}$~A. Finally, each individual $I$--$V$ curve was independently min-max normalized to a value range between 0 and 1. This normalization ensured that clustering was primarily governed by functional curve shape rather than absolute current magnitude.

Clustering was performed using the MiniSom implementation of SOMs \cite{Vettigli2018}. Each SOM consists of $n_\mathrm{neurons}$ neurons arranged as an intrinsically two-dimensional map within the 203-dimensional feature space of the processed current data (excluding data points for $V\in[-0.1,0.1]$~V as introduced in the preceding paragraph). Both rectangular and hexagonal map topologies as well as Euclidean and cosine distance metrics were evaluated. During training, each neuron competitively adapts toward the data distribution over $t$ iterations according to a learning rate $\eta(t)$ and a Gaussian neighborhood function with width $\sigma(t)$, thereby preserving local topological relationships within the reduced representation.

Hyperparameters (HPs) were optimized using five consecutive 5-fold cross-validated grid searches with iterative refinement of the search space. The explored parameter ranges and final HP set are summarized in Table~\ref{table:SOM_params}. The final model converged to $n_\mathrm{neurons}=1156$, consistent with the commonly used heuristic $n_\mathrm{neurons}\approx5\sqrt{n_\mathrm{data}}$ for SOM-based representations. Lower training iterations (e.g., $5\times10^4$) produced equivalent clustering results, indicating convergence of the trained representation.

\begin{table}
\caption{Search ranges and selected values of the SOM HPs. The number of neurons, initial learning rate, initial spread of the Gaussian neighborhood function, and number of iterations are denoted by $n_\mathrm{neurons}$, $\eta$, $\sigma$, and $t$, respectively. Topology and distance metrics are abbreviated as rectangular (Rect), hexagonal (Hex), cosine (Cos), and Euclidean (Euclid).
}
\label{table:SOM_params}
\begin{center}
\begin{tabular}{l c c}
\hline
HP & search range & selection\\
\hline
$n_\mathrm{neurons}$ & [9,5776] & 1156 \\
$\eta$ & [0.01,10] & 0.01 \\
$\sigma$ & [1,100] & 10 \\
$t$ & [$10^3$,$10^5$] & $10^5$ \\
topology & \{Rect,Hex\} & Rect \\
distance & \{Cos,Euclid\} & Euclid \\
\hline
\end{tabular}
\end{center}
\end{table}

Finally, the resulting 1156 SOM clusters (neurons) were aggregated into seven higher-level functional regime classes. This aggregation exploited the topology-preserving ordering of the SOM representation and was guided by feature engineering. The resulting functional regimes are discussed in detail in Section~\ref{sec:results}.

\subsubsection*{Cluster yield estimation}
\label{sssec:method_ML_cluster_yield}

The experimentally explored process space was sampled highly non-uniformly during approximately three years of empirical process optimization. To compensate for the resulting sampling bias and obtain statistically comparable cluster occupancies, a kernel-density-estimation (KDE)-based homogenization approach was applied.

For this purpose, the multidimensional process parameter space $\mathbf{x} \in \mathbb{R}^8$ was represented using KDEs, implemented in scikit-learn \cite{Parzen1962, Rosenblatt1956, Pedregosa2011}. One KDE was constructed using the complete dataset, while additional KDEs were generated individually for each of the seven functional clusters introduced in the preceding section. The relative ratio between cluster-specific ($P: \mathbb{R}^8 \to \mathbb{R}^7$) and total ($P_\mathrm{total}: \mathbb{R}^8 \to \mathbb{R}$) sampling densities defines the corresponding device yield $Y_\mathrm{device}(\mathbf{x})=P(\mathbf{x})/P_\mathrm{total}(\mathbf{x})$, independent of direct sampling density variations. Remaining indirect sampling effects are therefore limited primarily to local resolution accuracy.

\begin{table}
\caption{Search range defined either by the respective interval or complete set, and finally selected HP for the KDE. The kernel functions are abbreviated as follows Gaussian (Gaus), Exponential (Exp), and Linear (Lin). Please note that the HP values have been normalized to a range of 0--1 using min/max normalization.  Hence, all BWs are in arbitrary unit (a.u.).
}
\label{table:KDE_params}
\begin{center}
\begin{tabular}{l c c}
\hline
HP & search range & selection\\
\hline
kernel function & \{Gaus,Exp,Lin\} &  Exp\\
BW of $A_{\mathrm{device}}$ & [$10^{-5}$,$10^{3}$] & 0.26\\
BW of $F_\mathrm{Ar}/F_\mathrm{O_2}$ & [$10^{-5}$,$10^{3}$] & 5.11\\
BW of $P_{\mathrm{SiO}_x}$ & [$10^{-3}$,$10^{1}$] & 0.16\\
BW of $t_{\mathrm{SiO}_x}$ & [$10^{-5}$,$10^{3}$] & 0.10\\
BW of $r_\mathrm{wafer}$  & [$10^{-5}$,$10^{3}$] & 0.05\\
BW of $t_{\mathrm{measure}}$ & [$10^{-5}$,$10^{3}$] & 0.14\\
BW of $U_\mathrm{sweep}$ & [$10^{-5}$,$10^{3}$] & 0.16\\
BW of $t_{\mathrm{deg}}$ & [$10^{-5}$,$10^{3}$] & 0.10\\
\end{tabular}
\end{center}
\end{table}

KDE HPs, including kernel function and bandwidths for each process variable, were optimized using consecutive cross validated grid searches minimizing the root-mean-square deviation of predicted yields on previously unseen validation sets. The explored parameter ranges and final HP selection are summarized in Table~\ref{table:KDE_params}. Prior to KDE generation, all process parameters were independently min-max normalized to a value range between 0 and 1. Consequently, the optimized bandwidths directly provide information regarding the characteristic sensitivity scales of the corresponding process variables. Small bandwidths, such as for the radial wafer position $r_\mathrm{wafer}$ (BW $\ll1$), indicate rapidly varying functionality, whereas large bandwidths, such as for the process gas ratio $F_\mathrm{Ar}/F_\mathrm{O_2}$ (BW $>1$), indicate relative invariance within the experimentally explored process window.

The optimized bandwidths additionally define suitable discretization densities for evaluating the resulting multidimensional yield distribution $Y_\mathrm{device}(\mathbf{x})$, where $Y_\mathrm{device}: \mathbb{R}^8 \to \mathbb{R}^7$, for the process parameters $\mathbf{x}=
\left(
A_\mathrm{device},
F_\mathrm{Ar}/F_\mathrm{O_2},
P_{\mathrm{SiO}_x},
t_{\mathrm{SiO}_x},
r_\mathrm{wafer},
t_\mathrm{measure},
U_\mathrm{sweep},
t_\mathrm{deg}
\right)
$
using a sampling density proportional to $2\lceil1/\mathrm{BW}\rceil$ along each normalized process axis. The prefactor of 2 follows the Nyquist-Shannon sampling criterion \cite{Nyquist1928, Shannon1949}.

\begin{figure}
\includegraphics[width=8cm]{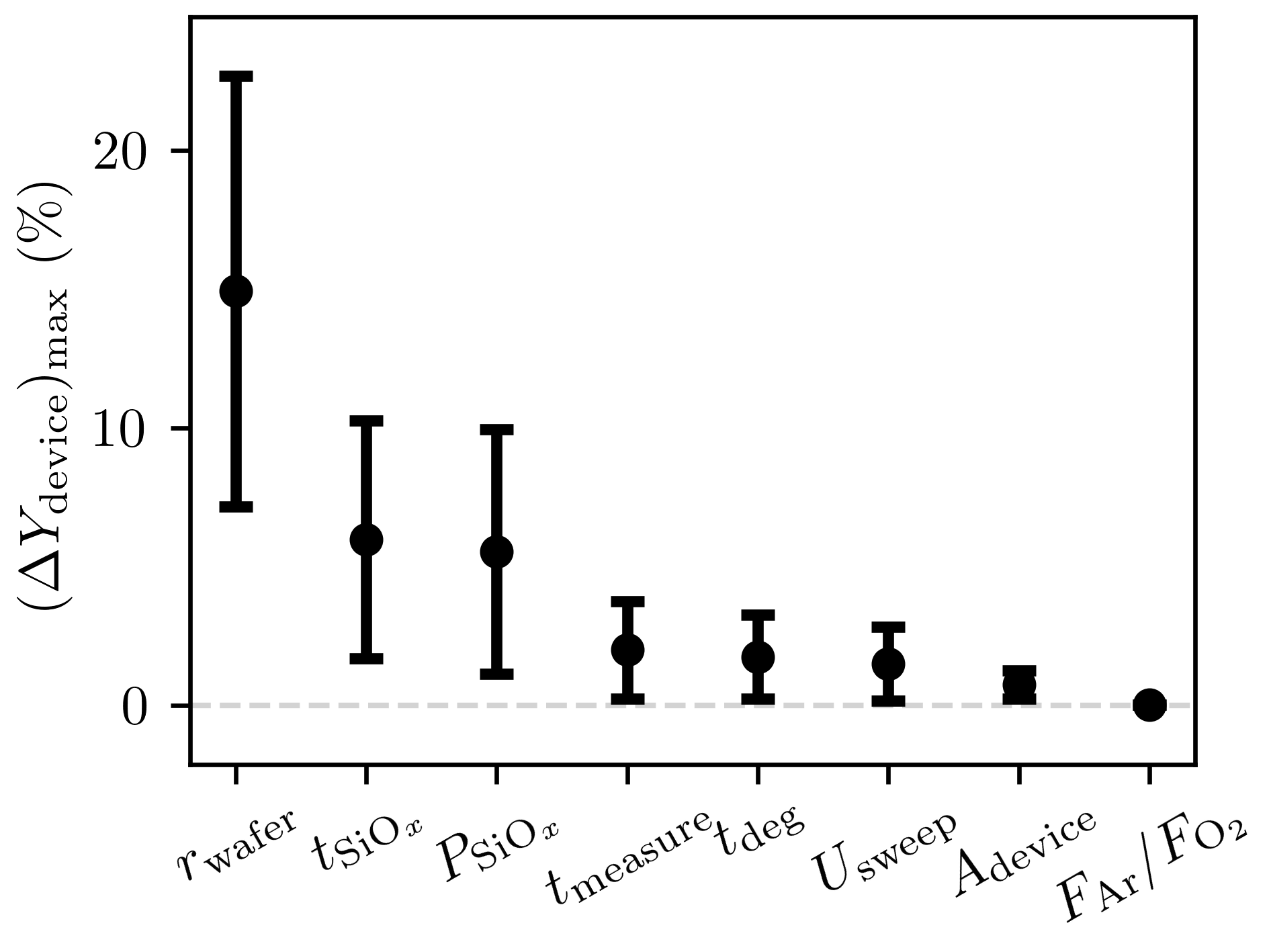}
\caption{Maximum change in device yield $(\Delta Y_\mathrm{device})_\mathrm{max}$ as a function of individual process parameters (marginalized accordingly). The markers' centers and ranges indicate the mean and root mean square deviations across the seven clusters, respectively.}
\label{fig:sensitivity}
\end{figure}

To identify the dominant process variables governing functional regime occupation, a sensitivity analysis was performed based on the maximum yield variation $\Delta(Y_\mathrm{device})_\mathrm{max}$ induced by each process variable independently, while marginalizing over all remaining dimensions. The results of this analysis are summarized in Figure~\ref{fig:sensitivity}. The markers indicate the mean maximum changes across the seven clusters, while the vertical ranges represent the corresponding root mean square deviation across clusters' mean values. Consistent with the optimized KDE bandwidths, only weak sensitivity with respect to $F_\mathrm{Ar}/F_\mathrm{O_2}$ was observed within the investigated process window. However, the weak observed sensitivity to O$_2$ fraction should be interpreted within the restricted poisoned-mode process window explored here (5-25~\% O$_2$). Prior studies have shown that substantially lower oxygen fractions, typically in the range of 0–1.25~\% O$_2$ near reactive transition boundaries, enable strong tuning of SiO$_x$ stoichiometry and, hence, device functionality (cf. Section~\ref{ssec:Plasma_and_atomistic}) \cite{Hattum2006, Tomozeiu2002}. Consequently, the analysis presented in Section~\ref{sec:results} focuses on the three dominant process variables $r_\mathrm{wafer}$, $t_{\mathrm{SiO}_x}$, and $P_{\mathrm{SiO}_x}$, while the remaining process dimensions were marginalized for the sake of simplicity and clarity.

\subsubsection*{Device-level simulations}
\label{sssec:methods_device_sim}

To reconstruct internal defect state characteristics underlying the experimentally observed switching behavior, device-level simulations were performed using a reduced one-dimensional (1D) cloud-in-a-cell (CIC) framework adapted from previous work on stochastic interface-based memristive systems \cite{Yarragolla2022}. 

In the following, the key aspects of device simulations relevant to the present study are outlined. These considerations form the basis for deriving physically plausible parameter sets that reproduce the different experimentally observed switching responses. Further details regarding the model formulation, numerical implementation, and parameterization, including the physical considerations linking Cu characteristics to TiN/SiO$_x$ Schottky barrier heights, oxygen-vacancy density to defect transport and electric-field redistribution, and defect-landscape dependent material properties, are provided in \cite{Yarragolla2026}.

The model captures electric-field-driven oxygen-vacancy drift coupled to Schottky and Poole--Frenkel electronic transport mechanisms, while additionally incorporating the influence of Cu through effective modulation of the TiN/SiO$_{x}$ Schottky barriers. Within this framework, the electrical response emerges from the coupled interaction between oxygen-vacancy distributions and dynamics, copper related defect landscape, and local electric field redistribution across the device. Importantly, the model intentionally represents the internal material state through effective defect-state coordinates only, without explicitly incorporating detailed chemical composition, atomistic network topology, or mechanical and structural descriptors. The reconstructed oxygen-vacancy and copper related state variables should therefore be interpreted as effective latent descriptors of the internal transport environment rather than direct physical concentrations.

\subsection{Forward engineering}
\label{ssec:method_forward}

The forward engineering workflow aims to identify the primary physical mechanisms governing defect-state formation as a function of plasma process conditions. Plasma and atomistic simulations were combined to establish causal links between deposition conditions, local Si-O network reconstruction, and resulting defect-state characteristics across scales.

\subsubsection*{Plasma simulations}
\label{sssec:method_simulation_plasma}

\begin{figure}
\includegraphics[width=8cm]{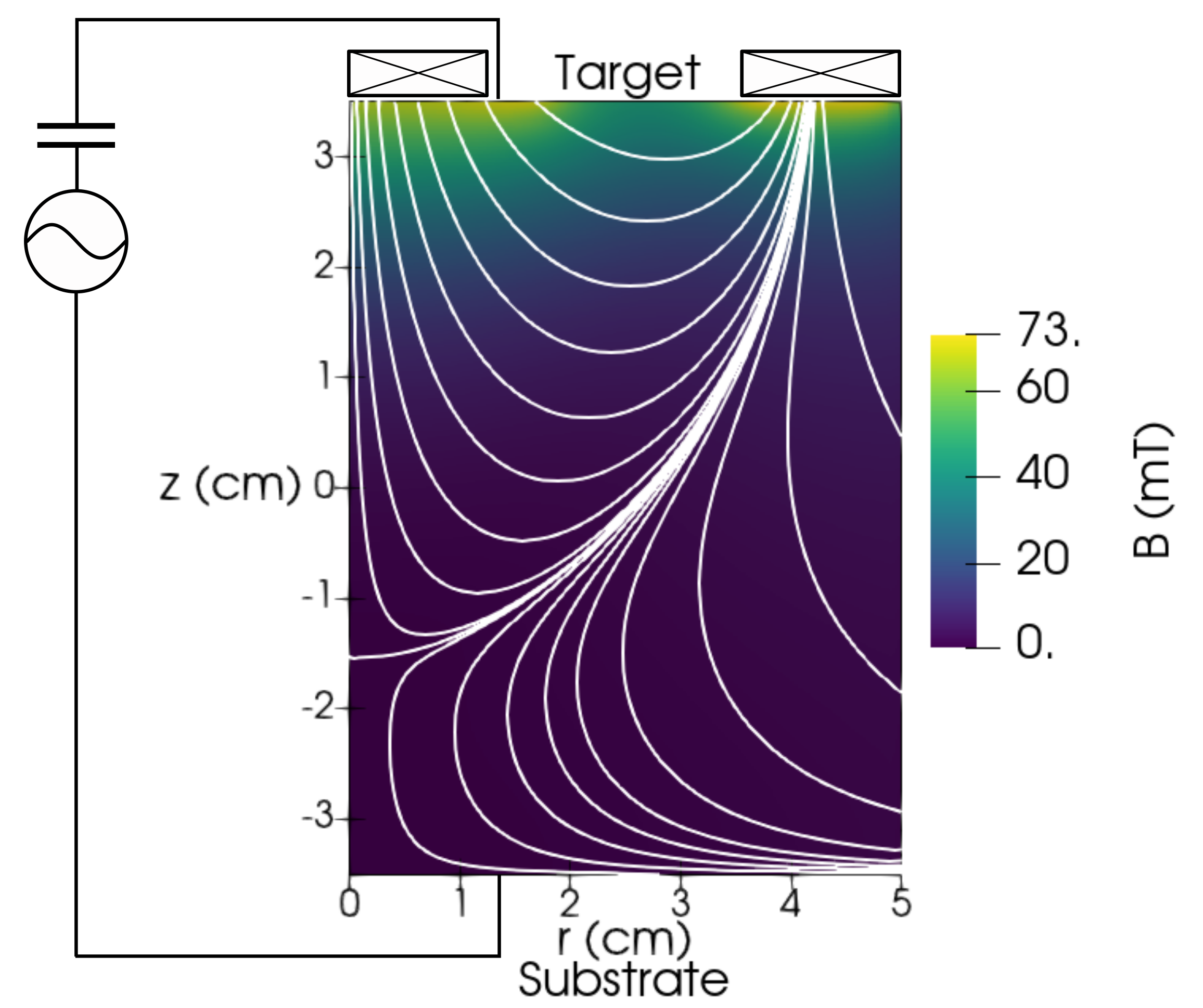}
\caption{Schematic of the cylindrically symmetric simulation domain and the magnetic field structure of the Ar/O\textsubscript{2} sputter deposition. Magnets are indicated above the target.}
\label{fig:plasma_geometry}
\end{figure}

Regarding a theoretical forward prediction of the device processing steps, we initially focus on the reactive sputter deposition of a-SiO$_x$. As discussed in Section~\ref{sec:closure}, this process step is most influential linking process to device parameters. To characterize the principle discharge physics, the discharge was simulated using cylindrically symmetric 2D/3V particle-in-cell/Monte Carlo collision (PIC-MCC) \cite{birdsallPlasmaPhysicsComputer1991}. A schematic of the simulation domain is depicted in Fig.~\ref{fig:plasma_geometry}. The target electrode with circular magnets on and connected through a blocking capacitor was driven by an radio-frequency (rf) voltage $V_\mathrm{rf}$. The outer wall and the substrate electrode were grounded. An Ar/O\textsubscript{2} gas phase chemistry was adopted \cite{guerraKineticModelLowpressure1999,babaevaOxygenIonEnergy2005,bultinckParticleincellMonteCarlo2009a,kutasiTheoreticalInsightAr2010a,gudmundssonBenchmarkStudyCapacitively2013,hannesdottirRoleMetastableO2b1Sigmag2016,derzsiExperimentalSimulationStudy2016,biagiBiagiDatabaseWwwlxcatnet2026,morganMorganDatabaseWwwlxcatnet2026}, whereas the species Ar$^+$, Ar($1\mathrm{s}_5$), Ar($1\mathrm{s}_4$), Ar($1\mathrm{s}_3$), Ar($1\mathrm{s}_2$), O$_2^+$, O$_2$(a$^1 \Delta_g$), O$_2$(b$^1 \Sigma_g^+$), O, O$^-$, Si, and electrons were dynamically evolved (Paschen notation). All charged species as well as sputtered Si and O were simulated kinetically. In contrast, a hybrid reaction-diffusion continuum description with time-slicing was used to simulate the neutral transport of Ar($1\mathrm{s}_5$), Ar($1\mathrm{s}_4$), Ar($1\mathrm{s}_3$), Ar($1\mathrm{s}_2$), O$_2$(a$^1 \Delta_g$), O$_2$(b$^1 \Sigma_g^+$), and O (not sputtered) \cite{laricchiutaClassicalTransportCollision2007,laricchiutaHighTemperatureMars2009}. 
O$_2$ Herzberg states, the O($^1$D) electronic excited state, and the O$_2$(X,$v$) vibrational manifold, which have been found to be non-negligible in DC glow discharges \cite{diasReactionMechanismOxygen2023}, are ignored here. This is justified by the low operating gas pressure and low O$_2$ admixture, whereas the O$_2$ density is (at least) 3 order of magnitude lower than in \cite{diasReactionMechanismOxygen2023}.
The evolution of the surface subject to reactive particle fluxes was simulated assuming effective chemical surface coverage fractions (sum unity) with surface sites s-Si, s-SiO, and s-SiO\textsubscript{2}. These were evolved taking into account the change corresponding to the incoming reactive fluxes and again accelerated using time-slicing \cite{Berg2005,tonneauUnderstandingRoleEnergetic2021}. Hence, a consistent steady-state prediction of the discharge and the surface conditions was obtained. Effects such as the poisoning of surfaces (in particular the target) were considered self-consistently with a direct influence on plasma--surface interaction and particle emission yields. Electron- and ion-induced secondary electron emission (SEE) were considered through analytical fitting formulae, linearly approximating their change due to varying surface conditions \cite{lyeTheorySecondaryEmission1957,youngDissipationEnergy25101957,toliasSecondaryElectronEmission2014,hannesdottirRoleMetastableO2b1Sigmag2016,phelpsColdcathodeDischargesBreakdown1999}. The ion energy-dependent sputtering of the varying SiO$_x$ target surface with surface chemical composition $x$, was predicted by a machine learning (ML) surrogate model. This data-driven model was adopted from previous works on sputtering TiAl composite targets and trained with data of Ar$^+$ and O$_2^+$ sputtering SiO$_x$ with varying ion energy, ion flux contributions, and surface chemical composition, obtained from Tridyn simulations \cite{Gergs2021, Möller1984, pruferComputerModelingSinglelayer2019}. The hybrid PIC-MCC/ML scheme was implemented in the OpenFOAM framework (foundation release, version 12) \cite{Weller1998, Weller2024}. The Ar/O$_2$ plasma simulation to be discussed was carried out at 5 \% O$_2$, 0.5~Pa, 300~K, $V_\mathrm{rf}=250$~V (amplitude).

\subsubsection*{Plasma-surface simulations} 
\label{sssec:method_simulation_PSI}

In this section, the methodological approach used to investigate plasma--surface interactions at the substrate during SiO$_x$ deposition is described. To isolate the intrinsic response of the SiO$_x$ network to the plasma-defined growth conditions, effects associated with the underlying layers are not explicitly considered in the forward simulations, i.e., TiN during deposition of the first SiO$_x$ layer and Cu islands during deposition of the second SiO$_x$ layer. Their influence on the reconstructed device state is, however, implicitly contained in the reverse-engineering approach, which derives the defect landscape from experimental data, as outlined in Section~\ref{ssec:method_reverse}.

The transient thermal response associated with energetic electron--surface interaction was estimated using a simplified continuum heat-transport model based on a two-dimensional finite-difference formulation with composition-dependent material properties. Heat transport was solved using an implicit Crank--Nicolson scheme. Further implementation details are provided in the Supplementary Information.

All ion--surface interactions were studied using the Large-scale Atomic/Molecular Massively Parallel Simulator (LAMMPS) \cite{Thompson2022}. Atomistic model systems spanning the stoichiometry range $x$=1.0--2.0 were generated using reactive MD based on the ReaxFF potential \cite{van_Duin2001, Nayir2019}. Initially, a reference a-SiO$_2$ structure was generated using an established repeated melting-and-quenching procedure described and validated elsewhere \cite{Fogarty2010, Nayir2019, Gergs2022}. The resulting bulk system consisted of 1150~Si and 2300~O atoms within a simulation cell of $3.65\times3.75\times3.75$~nm$^3$. Sub-stoichiometric systems were subsequently generated by sequential oxygen removal followed by structural relaxation. Starting from a-SiO$_2$, oxygen atoms were randomly removed to obtain the target stoichiometry, followed by isotropic relaxation at 1500~K and 0~Pa for 100~ps using a Berendsen thermostat ($\tau_\mathrm{damp}=100$~fs) and barostat ($\tau_\mathrm{damp}=1$~ps, $B=40$~GPa) \cite{Thompson2022, Berendsen1984}. The systems were subsequently cooled to 300~K using a previously validated cooling rate of 10~K/ps and equilibrated for an additional 100~ps after velocity re-initialization to remove processing artifacts \cite{Gergs2022}. This procedure was repeated sequentially in stoichiometry increments of $\Delta x=0.1$, where each relaxed configuration served as the initial structure for the subsequent lower-stoichiometry system.

Ion bombardment under sputter deposition conditions was simulated using a hybrid bulk/surface MD setup. Within this approach, the amorphous SiO$_x$ network was treated as a periodic bulk system, while incoming Ar ions interacted with a localized surface-like region. The applicability of this hybrid representation to the investigated ion-energy regime is discussed in the Supporting Information. For each stoichiometry, Ar bombardment simulations were performed for ion-to-Si flux ratios ranging from 0 to 1. Ar ions were inserted at random lateral positions with an impact energy of approximately 35~eV, based on the plasma simulations introduced in the preceding section. Ion interactions with the SiO$_x$ matrix were described using a hybrid ReaxFF/ZBL potential \cite{van_Duin2001, Nayir2019, Ziegler1985}, where short-range ZBL interactions captured collision dynamics while reactive interactions governed defect formation and structural relaxation.

After insertion, each Ar ion was propagated for 1~ps to resolve impact dynamics and local collision processes. During this stage, the impact region (defined as half of the simulation cell along the impact direction) was propagated without additional damping to avoid perturbation of the collision dynamics, whereas regions outside the local impact zone were weakly coupled to a Langevin thermostat to dissipate excess heat and suppress artificial energy accumulation \cite{Thompson2022, Brünger1984}. Subsequently, the entire SiO$_x$ system was relaxed for an additional 1~ps using a Langevin thermostat ($\tau_\mathrm{damp}=100$~fs) at 300~K together with a Berendsen barostat ($\tau_\mathrm{damp}=1$~ps, $B=40$~GPa) acting along the impact direction \cite{Thompson2022, Berendsen1984}. This thermostat and barostat were chosen to provide efficient thermal and mechanical relaxation rather than to rigorously sample an isothermal--isobaric equilibrium ensemble, as the simulations describe intrinsically nonequilibrium impact and relaxation dynamics. Ar atoms identified as reflected or outgassed during the simulation were removed, whereas retained Ar atoms remained within the simulation domain. This impact-relaxation sequence was repeated until target ion-to-Si flux ratios between 0 and 1 were reached.

The descriptors used for the main analysis comprised oxygen-vacancy density, dangling-Si density, non-bridging-oxygen density, mass density, stress, Si-O bond density, and Si-Si bond density. Si-O and Si-Si bonds were identified using distance cutoffs of 1.8~\AA{} and 2.7~\AA{}, respectively. The Si coordination number was defined as the total number of Si-O and Si-Si neighbors within these cutoffs. Dangling Si defects were identified as under-coordinated Si atoms with coordination numbers below four, while non-bridging oxygen defects were defined as O atoms bonded to one or fewer Si neighbors. Oxygen vacancy-like defects were identified using a more restrictive geometric criterion due to the pronounced sensitivity of simple Si-Si distance-based definitions. Candidate Si pairs were first selected from under-coordinated Si atoms separated by distances between the Si-Si bond cutoff and a vacancy cutoff of 3.2~\AA{}. For each candidate pair, possible missing-oxygen bonding directions were evaluated using cone sampling around the Si-Si axis. A candidate vacancy was accepted only if compatible bonding directions were available at both Si atoms and if the corresponding region was not sterically blocked by neighboring Si, O, or Ar atoms. Cone blocking was evaluated using a cosine cutoff of 0.6, a steric exclusion radius of 1.6~\AA{}, and angular sampling within a maximum cone angle of $\pi/6$. Stress was obtained from the time-averaged pressure tensor as $-(p_{xx}+p_{yy}+p_{zz})/3$. Owing to the continuous stress relaxation along the out-of-plane direction ($p_{zz}\approx0$), this quantity is directly proportional to the in-plane biaxial film stress commonly reported for sputter-deposited thin films and is therefore used as its atomistic proxy throughout this work. Defect and bond densities were normalized by the instantaneous simulation volume and reported in nm$^{-3}$. To suppress stochastic fluctuations originating from individual impact events while preserving underlying trends, the resulting property maps were post-processed using weak two-dimensional Gaussian smoothing. The kernel width corresponded to one stoichiometry interval ($\sigma_x\approx0.1$ in $x$) and five sampling intervals (out of 41) along the logarithmically sampled ion-to-Si flux ratio axis.

The robustness of the structural descriptors with respect to the short impact-relaxation sequence was verified by additional equilibration of selected configurations (see Supporting Information). The resulting structural properties remained largely unaffected, whereas larger deviations were observed for stress-related quantities (see Supporting Information); consequently, the higher-resolution impact-series configurations were used to construct the structural maps presented in Section~\ref{sec:results}.

\section*{Acknowledgement}

Funded by the Deutsche Forschungsgemeinschaft (DFG, German Research Foundation) in the frame of SFB 1461 (Project-ID 434434223), Research Grant MU 2332/18-1 (Project-ID 546680029), and Research Grant TR 1625/1-1 (Project-ID 568560111).

\section*{Data Availability}

The data that support the findings of this study are available from the corresponding author upon reasonable request.


\section*{ORCID}
\noindent
T. Gergs: \url{https://orcid.org/0000-0001-5041-2941} \\
R. Lamprecht \url{https://orcid.org/0009-0000-2324-6714} \\
S. Yarragolla \url{https://orcid.org/0000-0002-2973-4943} \\
O. Gronenberg \url{https://orcid.org/0009-0003-0232-1078} \\
L. Vialetto \url{https://orcid.org/0000-0003-3802-8001} \\
H. Kohlstedt \url{https://orcid.org/0000-0002-5181-8633} \\
T. Mussenbrock: \url{https://orcid.org/0000-0001-6445-4990} \\
J. Trieschmann: \url{https://orcid.org/0000-0001-9136-8019}

\clearpage

\bibliography{./references.bib}


\clearpage


\section*{Supplementary material}

# Supplementary Information

# From Processing to Functionality: Engineering Accessible Material States in Cu-Embedded $SiO_x$ Memristive Devices

Tobias Gergs,[1,2,3,*] Rouven Lamprecht,[4] Sahitya Yarragolla,[1,2,5] Ole Gronenberg,[5,6] Luca Vialetto,[1,7] Hermann Kohlstedt,[4,5] Thomas Mussenbrock,[2] and Jan Trieschmann[1,5,†]

[1] *Theoretical Electrical Engineering, Department of Electrical and Information Engineering, Kiel University, Kaiserstraße 2, 24143 Kiel, Germany*

[2] *Chair of Applied Electrodynamics and Plasma Technology, Department of Electrical Engineering and Information Science, Ruhr University Bochum, 44780 Bochum, Germany*

[3] *Fraunhofer Institute for Electronic Nano Systems ENAS, 09126 Chemnitz, Germany*

[4] *Nanoelectronics, Department of Electrical and Information Engineering, Kiel University, Kaiserstraße 2, 24143 Kiel, Germany*

[5] *Kiel Nano, Surface and Interface Science KiNSIS, Kiel University, Christian-Albrechts-Platz 4, 24118 Kiel, Germany*

[6] *Synthesis and Real Structure, Department of Materials Science, Kiel University, Kaiserstraße 2, 24143 Kiel, Germany*

[7] *Department of Mechanical and Aerospace Engineering, University of California, Los Angeles, Los Angeles, CA 90095, United States of America*



## S1. RESULTS

### S1.1. Emergent switching regimes and functional state space

The enlarged depiction of the U-matrix in Figure S1 is primarily intended to provide a clearer visual representation of the $I$–$V$ characteristic curves for each neutron (cluster), compared to the compact version in the main paper.

* tobias.gergs@enas.fraunhofer.de

† jt@tf.uni-kiel.de

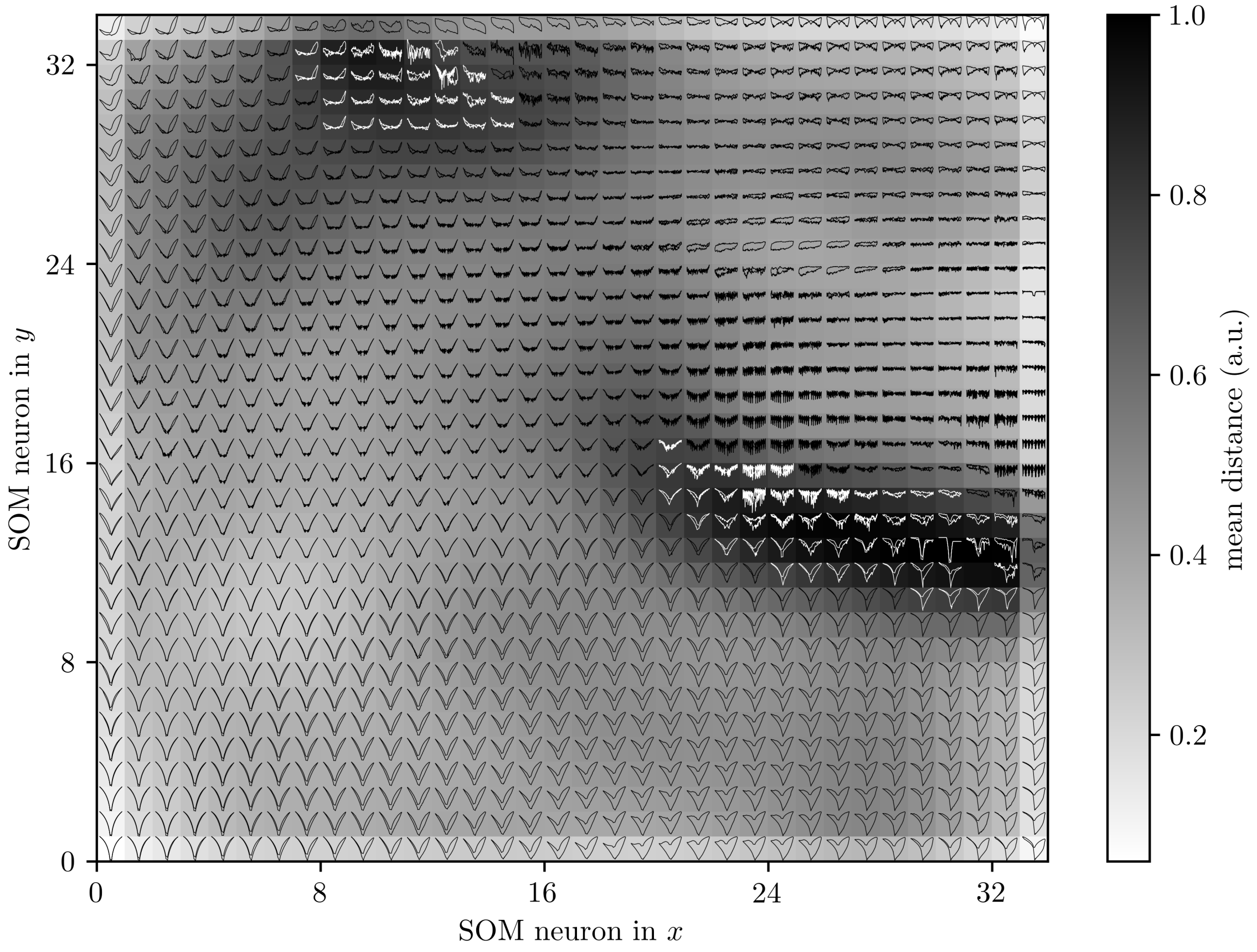


Figure S1. Enlarged U-matrix predicted by SOM, superimposed with the $I$–$V$ curves centroids of the respective 34 × 34 clusters.

### S1.2. Physics-informed inference of latent defect state landscapes

Representative wafer maps are shown in Figure S2 to illustrate the spatial variability of the functional cluster distributions both within individual wafers and across wafers fabricated under similar or different process conditions. The gray disks indicates the total wafer area. Since some wafers were processed off-center with respect to the symmetry axis of the plasma chamber, the corresponding distributions were re-centered during postprocessing. For this purpose, the cluster distributions were represented in polar coordinates, and the symmetry center was determined by minimizing their azimuthal variance, consistent with the expected radial symmetry of the processing chamber. The resulting symmetry centers, indicated by black crosses in Figure S2, were then used to re-center the device coordinates $(x, y)$ and derive a consistent radial device position $r_{\text{wafer}}$ relative to the processing symmetry axis for

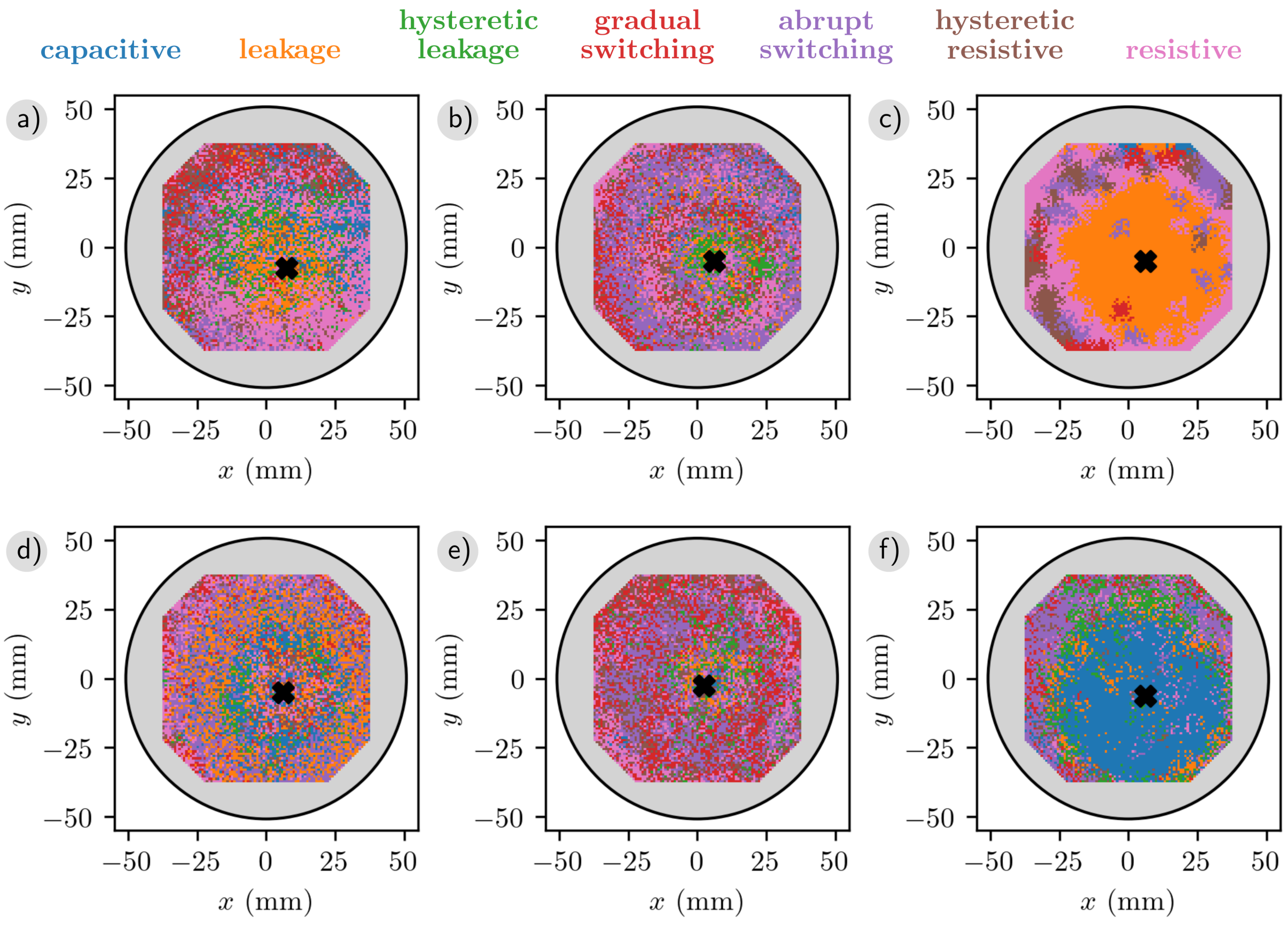


Figure S2. Examples of cluster distributions on wafers. The wafers are indicated as gray disks. Off-center processed wafers have been re-centered in the postprocessing. The obtained plasma processing chamber's symmetry centers are marked by black crosses. $x$ and $y$ are the positions on (off-centric) the wafer. Power $P_{\mathrm{SiO}_x}$ and deposition time $t_{\mathrm{SiO}_x}$ were a) 125 W and 50 s, b) 125 W and 100 s, c) 125 W and 50 s, d) 170 W and 40 s, e) 200 W and 32 s, f) 200 W and 62 s, respectively.

all wafers. The resulting $r_{\mathrm{wafer}}$ values were subsequently used as the wafer-position process parameter for evaluating the process-dependent cluster yields in the main manuscript. In other words, wafers processed off-center were virtually shifted along $x$ and $y$ to establish a common spatial reference frame across all wafers.

Figure S3 shows the reconstructed effective Cu density. As explained in the main paper, the Cu density is a function of the oxygen vacancy density; thus, while this information is technically redundant, it may nonetheless be relevant to the interested reader.

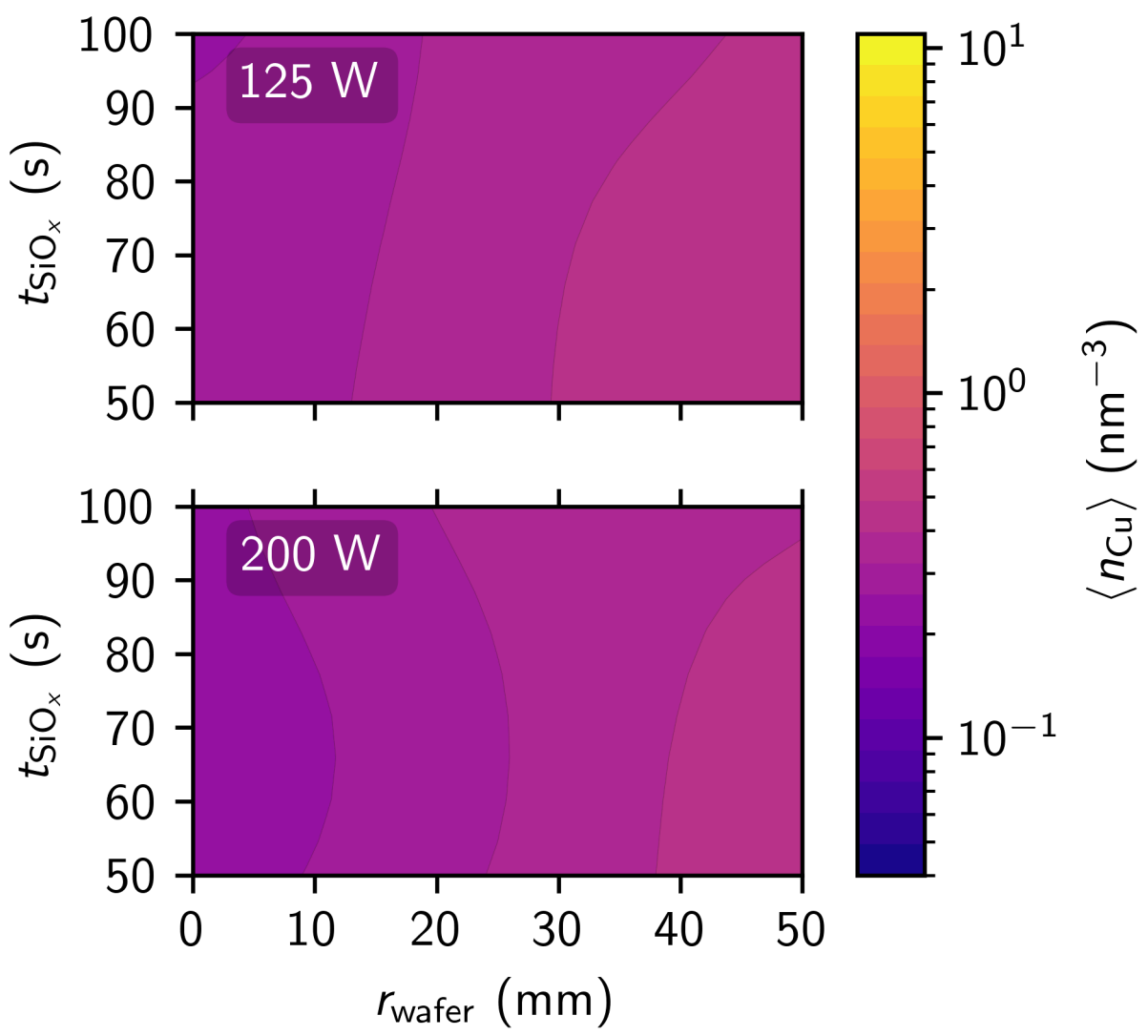


Figure S3. The expected value of the dissolved copper densities as a function of the radial device site on the wafer $r_{\mathrm{wafer}}$ and deposition time $t_{\mathrm{SiO}_x}$. The power $P_{\mathrm{SiO}_x}$ is included in each top left corner.

### S1.3. Plasma and atomistic origins of defect formation

The electron density and the corresponding electric potential obtained from the plasma simulations and averaged over 500 rf-cycles are shown in Figure S4. As electron transport is restricted within the magnetic field, electrons are confined to the ionization region where magnetic field lines are approximately parallel to the target surface. As expected, the electron density drops with distance from the cathode. A dc self-bias voltage of approximately -110 V establishes, whereas the peak voltage in the plasma bulk reaches up to approximately 30 V. Ions created in the ionization region are predominantly subject to acceleration toward the target. Ions created in the bulk region and subject to acceleration to the substrate surface are indirectly affected by the magnetic field due to the electric potential that builds up to ensure quasi-neutrality in the plasma bulk (in the presence of negative oxygen ions $O^-$).

The simulated transient temperature evolution is shown in Figure S5. Following the conservative assumption of complete local conversion of the electron kinetic energy into heat, the temperature rise remains highly localized and rapidly dissipates into the TiN bottom electrode. Depending on the $SiO_x$ stoichiometry, thermal relaxation occurs within approximately 1–10 ps. Since this represents an upper-bound estimate of the deposited thermal

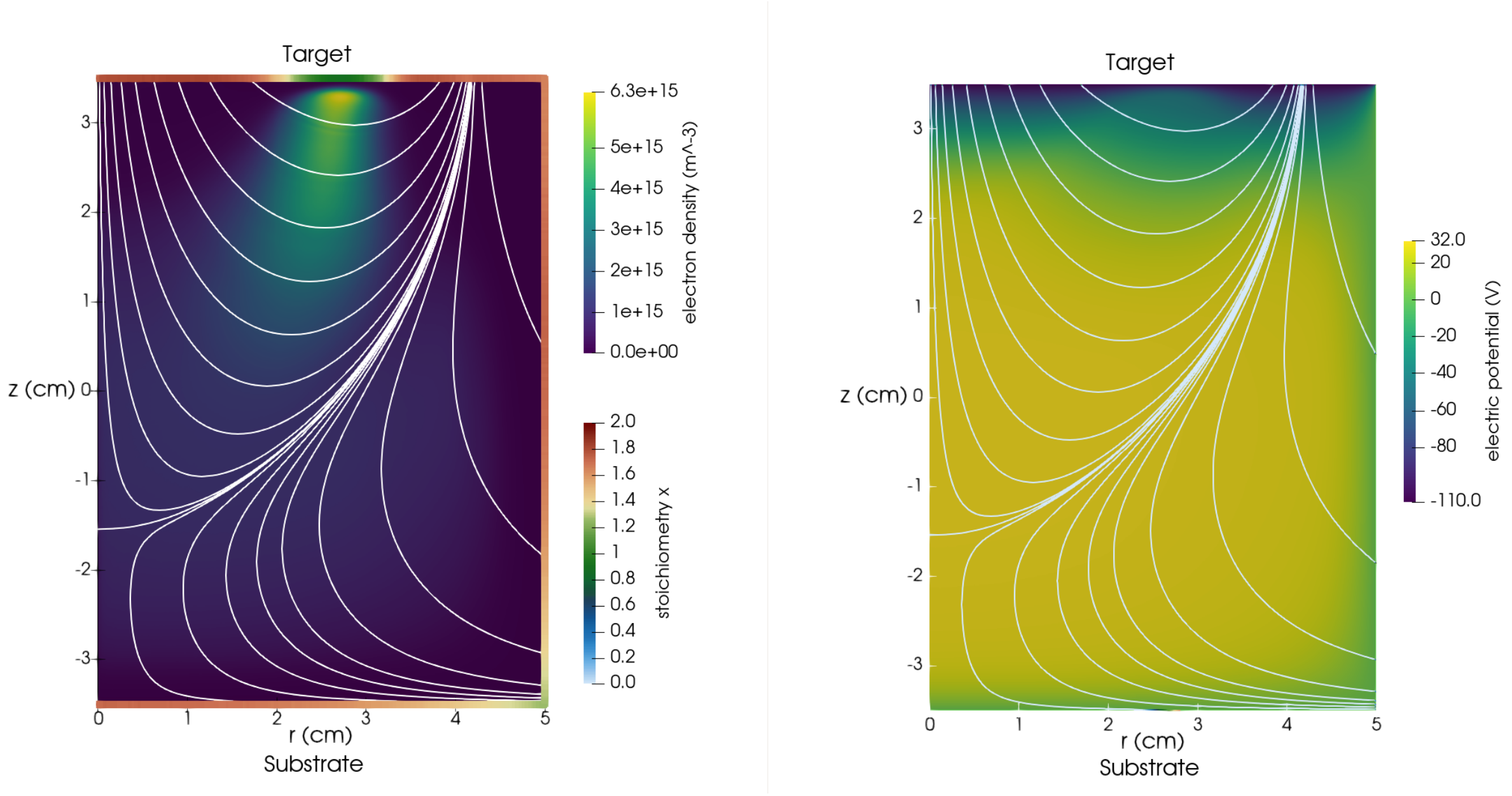


Figure S4. Cross-sections in the $rz$-plane of the average electron density (left) and average electric potential (right). The magnetic field structure and the effective stoichiometry of the surfaces are overlayed. Simulation parameters: $Ar/O_2$ with 5 % $O_2$, 0.5 Pa, 300 K, $V_{\mathrm{rf}} = 250$ V

energy, the actual thermal contribution of plasma electrons is expected to be even smaller. These results therefore suggest that electron-induced heating is unlikely to constitute the dominant origin of the experimentally observed structural modifications, although it may contribute to short-lived local relaxation processes.

To visualize the investigated material systems and their structural evolution under ion bombardment, representative atomistic configurations are shown in Figure S6. Silicon, oxygen, and argon atoms are rendered in yellow, red, and purple, respectively. All renderings were generated using OVITO [1].

Along the vertical direction (e.g., a→d→g), the stoichiometry decreases from a-$SiO_2$ to a-$SiO_{1.5}$ and a-$SiO_1$, respectively. The reduced oxygen content together with the increasing mass density (see main text) results in noticeably smaller simulation cell volumes.

Along the horizontal direction (e.g., a→b→c), ion-to-Si flux ratios $\Gamma_{\mathrm{Ar}^+}/\Gamma_{\mathrm{Si}}$ of 0, 0.1, and 1 are considered. The most apparent structural change is the progressive incorporation of Ar atoms into the amorphous network (see main text). In addition, increasing ion bombardment induces a progressively stronger relaxation of the simulation cell along the surface-normal direction, reflecting the accumulation and subsequent relaxation of compres-

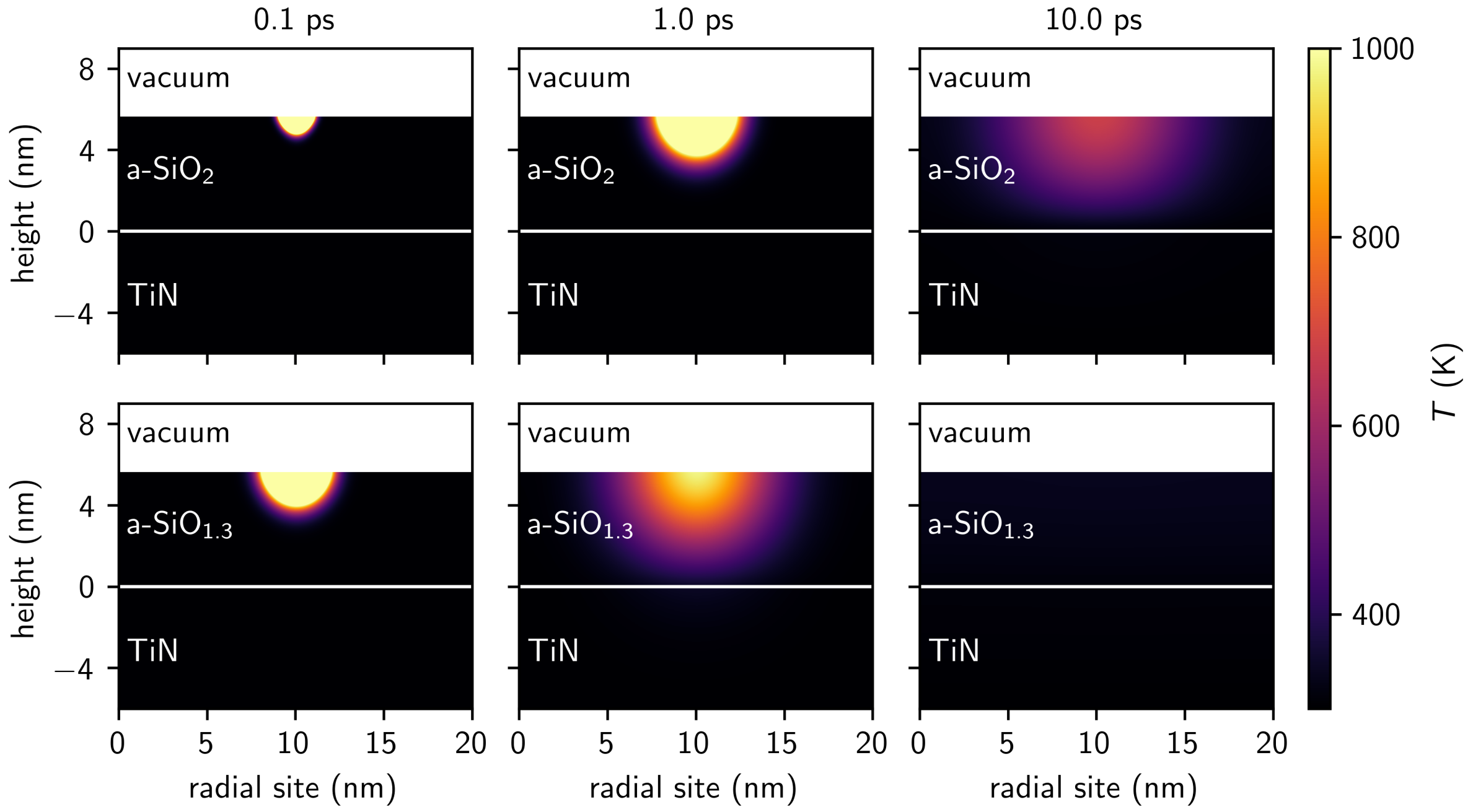


Figure S5. Simulated transient thermal response following a single energetic electron impact. Temperature evolution in representative 6 nm-thick a-$SiO_2$ (top) and a-$SiO_{1.3}$ (bottom) films following the instantaneous deposition of the complete kinetic energy of a 25 eV electron into a single surface cell. The latter represents an upper-bound estimate of the local thermal loading. Heat is dissipated into the TiN bottom electrode, resulting in rapid thermal relaxation.

sive growth stresses. This response is strongly stoichiometry dependent, with maximum out-of-plane expansions of 0.06, 0.05, and 0.12 nm for a-$SiO_2$, a-$SiO_{1.5}$, and a-$SiO_1$, respectively, consistent with the nonlinear mechanical response discussed in the main text.

Complementing the structural descriptors presented in the main paper, Figure S7 shows the corresponding evolution of the internal stress state. Note that the $\Gamma_{Ar^+}/\Gamma_{Si}$ axis is displayed on a linear rather than logarithmic scale. This representation enables the use of the additionally equilibrated configurations, which were sampled equidistantly in ion-to-Si flux ratio. In contrast, the logarithmically sampled impact-series configurations employed in the main paper are better suited for structural descriptors but exhibit substantially larger deviations for stress-related quantities (see main paper). For a-$SiO_2$, the compressive stress increases continuously with increasing ion bombardment. A qualitatively similar trend is observed for lower stoichiometries; however, the stress buildup becomes progressively weaker. In particular, for the structurally softened regime below $x \approx 1.7$, the stress saturates for $Ar^+$-

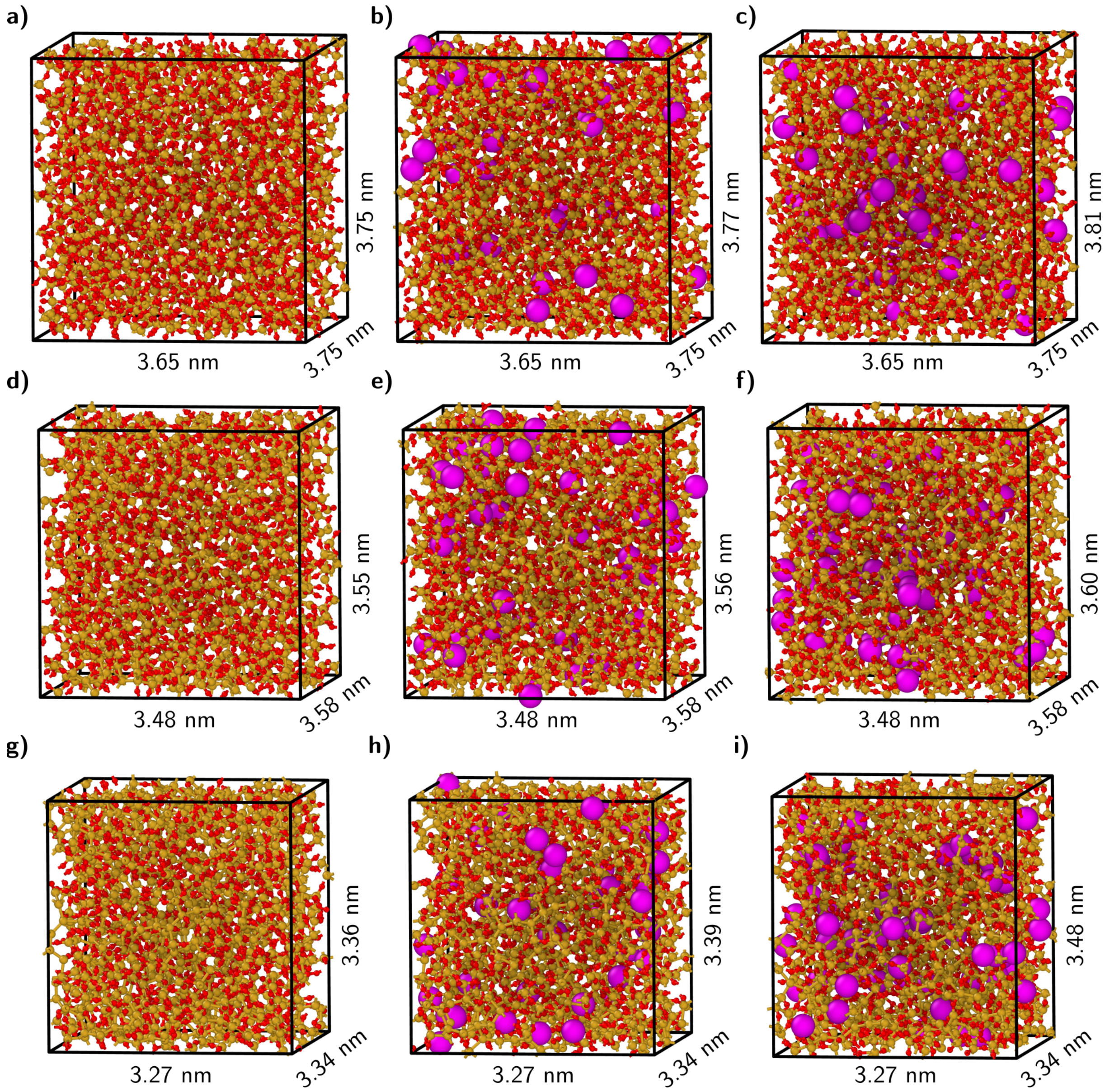


Figure S6. Visualization of atomic a-$SiO_x$ for various $Ar^+$ flux to $SiO_x$ flux ratios $\Gamma_{Ar+}/\Gamma_{Si}$. Along the vertical axis (top $\rightarrow$ bottom), the stoichiometry $x$ varies from 2 to 1.5 to 1, and along the horizontal axis (left $\rightarrow$ right), the flux ratio varies from 0 to 0.1 to 1. a) a-$SiO_2$, $\Gamma_{Ar+}/\Gamma_{SiO_x}$=0; b) a-$SiO_2$, $\Gamma_{Ar+}/\Gamma_{Si}$=0.1; c) a-$SiO_2$, $\Gamma_{Ar+}/\Gamma_{Si}$=1; d) a-$SiO_{1.5}$, $\Gamma_{Ar+}/\Gamma_{Si}$=0; e) a-$SiO_{1.5}$, $\Gamma_{Ar+}/\Gamma_{Si}$=0.1; f) a-$SiO_{1.5}$, $\Gamma_{Ar+}/\Gamma_{Si}$=1; g) a-SiO, $\Gamma_{Ar+}/\Gamma_{Si}$=0; h) a-SiO, $\Gamma_{Ar+}/\Gamma_{Si}$=0.1; i) a-SiO, $\Gamma_{Ar+}/\Gamma_{Si}$=1. All images were rendered using OVITO.

to-$SiO_x$ flux ratios $\Gamma_{Ar+}/\Gamma_{SiO_x} \gtrsim 10^{-2}$ instead of increasing further. As discussed in the main paper, this behavior reflects the increasing ability of the Si-rich network to accommodate

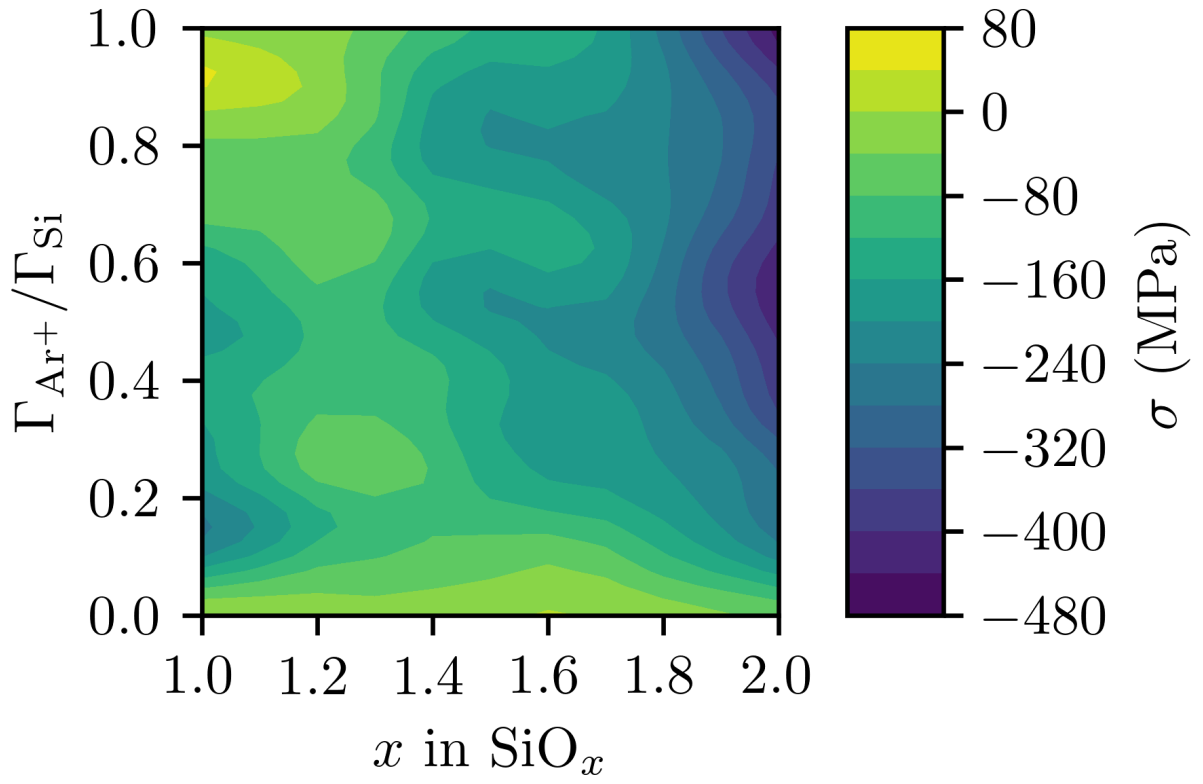


Figure S7. Results of Ar ion-$SiO_x$ interaction simulations for $x \in [0, 1]$ in $SiO_x$ and $Ar^+$ flux to $SiO_x$ flux ratio $\Gamma_{Ar^+}/\Gamma_{SiO_x} \in [0, 1]$ with mean ion energies of 35 eV. The stress $\sigma$ is shown. The $\Gamma_{Ar^+}/\Gamma_{Si}$ axis is displayed on a linear scale, unlike the logarithmic representation used for the structural descriptors in the main paper.

ion-induced deformation through topological network rearrangements, rather than continued elastic stress accumulation.

## S2. METHODS

### S2.1. Forward engineering

*Plasma-surface simulations*

To estimate the transient thermal response associated with energetic electrons, a simplified continuum heat-transport model was employed. The a-$SiO_x$/TiN stack was represented by a two-dimensional finite-difference model with composition-dependent thermal conductivity, specific heat capacity, and mass density. Heat transport was solved using an implicit Crank–Nicolson scheme with a fixed-temperature boundary at the TiN bottom electrode and thermally insulating boundaries elsewhere. The temperature field was obtained from the transient heat equation

$$\rho c_p \frac{\partial T}{\partial t} = \nabla \cdot (k \nabla T) \,, \tag{S1}$$

where $T$ denotes the temperature, $k$ the thermal conductivity, $\rho$ the mass density, and

$c_p$ the specific heat capacity. The deposited electron energy was introduced as the initial temperature field at $t = 0$. As a conservative upper-bound approximation, the complete kinetic energy of a single 25 eV plasma electron was assumed to be instantaneously converted into heat within a single surface cell. This intentionally overestimates the local temperature rise and therefore provides an upper bound for the thermal relaxation time.

Table S1. Material parameters used for the transient heat-transport simulations. Material properties of intermediate $SiO_x$ compositions were obtained by linear interpolation between a-$SiO_2$ and SiO.

| Property | a-$SiO_2$ | SiO | TiN |
|---|---|---|---|
| Thermal conductivity $k$ ($\mathrm{W\,m^{-1}\,K^{-1}}$) | 1.3 | 10 | 30 |
| Mass density $\rho$ ($\mathrm{kg\,m^{-3}}$) | 2200 | 2130 | 5220 |
| Specific heat capacity $c_p$ ($\mathrm{J\,kg^{-1}\,K^{-1}}$) | 750 | 710 | 550 |

Simulations were performed for representative 6 nm-thick a-$SiO_2$ and a-$SiO_{1.3}$ films to illustrate the influence of a-$SiO_x$ stoichiometry on the transient thermal response.

Ion–surface interaction was simulated using a hybrid bulk/surface molecular dynamics (MD) approach. The applicability of this hybrid bulk/surface representation is primarily supported by the comparatively low ion energies considered here. At impact energies of approximately 35 eV, energy dissipation remains strongly localized and predominantly induces intrinsic point-defect generation and local network rearrangements rather than extended collision cascades or large-scale surface reconstruction. Consequently, the resulting defect structures predominantly reflect localized defect formation and network evolution expected under experimentally relevant sputter-deposition conditions. While explicit growth simulations based on periodic atom insertion (i.e., every few picoseconds) in MD can in principle model film formation more directly, they introduce an artificially high particle flux that may distort the growth dynamics [2]; thus, both the static slab approach and growth simulations entail different, non-trivial methodological biases, and the choice between them depends on the physical regime of interest.

To assess the robustness of the short impact-relaxation sequence employed during ion bombardment, selected configurations covering all investigated $SiO_x$ stoichiometries ($\Delta x = 0.1$) and sampled at flux-ratio intervals of 0.025 were subjected to an additional equilibration procedure consisting of 90 ps relaxation at 300 K and 0 Pa along the impact direction with

Nose-Hoover style thermostat ($\tau_{\text{damp}} = 100$ fs) and barostat ($\tau_{\text{damp}} = 1$ ps) [3–7], followed by 10 ps of time averaging. Comparison between directly extracted impact-series configurations and additionally equilibrated configurations revealed only minor deviations for all structural descriptors considered in this work. Mean percentage deviations amounted to 3.22 %, 1.80 %, and 19.18 % for the Si dangling-bond, non-bridging oxygen, and oxygen-vacancy densities, respectively, as well as 0.01 % and 1.92 % for the Si-O and Si-Si bond densities. The mass density deviated by only 0.09 %. In contrast, stress-related quantities exhibited deviations of up to 166.56 %. Since stress serves only as a secondary descriptor in this work, the higher-resolution impact-series configurations were used to construct the structural maps presented in the main manuscript.

## ACKNOWLEDGEMENT

Funded by the Deutsche Forschungsgemeinschaft (DFG, German Research Foundation) – Project-ID 434434223 – SFB 1461 and Research Grant MU 2332/18-1 (Project-ID 546680029).

## DATA AVAILABILITY

The data that support the findings of this study are available from the corresponding author upon reasonable request.

## ORCID

T. Gergs: `https://orcid.org/0000-0001-5041-2941`
R. Lamprecht `https://orcid.org/0009-0000-2324-6714`
S. Yarragolla `https://orcid.org/0000-0002-2973-4943`
O. Gronenberg `https://orcid.org/0009-0003-0232-1078`
L. Vialetto `https://orcid.org/0000-0003-3802-8001`
H. Kohlstedt `https://orcid.org/0000-0002-5181-8633`
T. Mussenbrock: `https://orcid.org/0000-0001-6445-4990`
J. Trieschmann: `https://orcid.org/0000-0001-9136-8019`




\end{document}